\pdfoutput=1
\newcommand*{\ATLASLATEXPATH}{latex/}
\documentclass[cernpreprint, USenglish, texlive=2016]{\ATLASLATEXPATH atlasdoc}
\usepackage{\ATLASLATEXPATH atlaspackage}
\usepackage{\ATLASLATEXPATH atlasbiblatex}
\usepackage{\ATLASLATEXPATH atlasphysics}
\DeclareFieldFormat[article]{pages}{#1}%
\graphicspath{{logos/}{figures/}}
\usepackage{EXOT-2016-18-PAPER-defs}
\usepackage{multirow} 
\usepackage{graphicx}
\usepackage{siunitx}
\usepackage{calc}
\usepackage{changepage}

\newcommand*{\alignUnderEnDash}[1]{\multicolumn{1}{r@{}}{\llap{\makebox[\widthof{--}][c]{#1}}}}
  
\AtlasTitle{Search for vector-boson resonances decaying to a top quark and bottom quark in the lepton plus jets final state in $pp$ collisions at $\sqrt{s}$ = 13~\tev\ with the ATLAS detector}

\author{The ATLAS Collaboration}

\AtlasRefCode{EXOT-2016-18}

\PreprintIdNumber{CERN-EP-2018-142}

\AtlasJournal{Phys.\ Lett.\ B.}
\AtlasJournalRef{\PLB 788 (2019) 347}
\AtlasDOI{10.1016/j.physletb.2018.11.032}

\date{\today}

\AtlasAbstract{%
A search for new charged massive gauge bosons, \Wp, is performed with the ATLAS detector at the LHC. Data were collected in proton--proton collisions 
at a centre-of-mass energy of $\sqrt{s}$ = 13~\tev\ and correspond to an integrated luminosity of \lumi.
This analysis searches for \Wp~bosons in the $W^\prime  \rightarrow t\bar{b}$ decay channel in final states
with an electron or muon plus jets. 
The search covers resonance masses between 0.5 and 5.0~\tev\ and considers right-handed \Wp~bosons.
No significant deviation from the Standard Model (SM) expectation is observed and upper limits 
are set on the $W^\prime \rightarrow t\bar{b}$ cross section times branching ratio 
and the \Wp~boson effective couplings as a function of the \Wp~boson mass.
For right-handed \Wp~bosons with coupling to the SM particles equal to the SM weak coupling constant, masses below 3.15~\tev\ are excluded at the 95\% confidence level.
This search is also combined with a previously published ATLAS result for $W^\prime \rightarrow t\bar{b}$ in the fully hadronic final state.
Using the combined searches, right-handed \Wp~bosons with masses below 3.25~\tev\ are excluded at the 95\% confidence level.
}

\AtlasCoverSupportingNote{Search for $W^\prime \rightarrow t\bar{b}$ in the lepton+jets channel in $pp$ collisions}{https://cds.cern.ch/record/2223889}
\AtlasCoverCommentsDeadline{---}

\AtlasCoverAnalysisTeam{}

\AtlasCoverEdBoardMember{}

\hypersetup{pdftitle={ATLAS document},pdfauthor={The ATLAS Collaboration}}
\begin{document}
%-------------------------------------------------------------------------------
\maketitle
%-------------------------------------------------------------------------------
%\tableofcontents
%-------------------------------------------------------------------------------
% List of contributors - print here or after the Bibliography.
%\PrintAtlasContribute{0.30}
%\clearpage
%-------------------------------------------------------------------------------
\section{Introduction}
\label{sec:intro}
Many approaches to theories beyond the Standard Model (SM) introduce new charged vector currents mediated by heavy gauge bosons, usually referred to as \Wp. 
For example, the \Wp\ boson can appear in theories with universal extra dimensions, such as Kaluza--Klein excitations 
of the SM $W$ boson~\cite{Kaluza-Klein1,Kaluza-Klein2,Kaluza-Klein3},
or in models that extend fundamental symmetries of the SM and propose a massive right-handed counterpart to the 
$W$ boson~\cite{left-right-model1,left-right-model2,left-right-model3}.
Little-Higgs~\cite{little-Higgs} and composite-Higgs~\cite{Dugan:1984hq,Agashe:2004rs} theories also predict a \Wp\ boson.
The search for a \Wp\ boson decaying into a top quark and a $b$-quark (illustrated in Figure~\ref{fig:wprimeFeyn}) explores models
potentially inaccessible to searches for a \Wp\ boson decaying into leptons~\cite{Wprime_lnu0,Wprime_lnu1,CMS-EXO-12-060,EXOT-2013-10,EXOT-2015-06,CMS-EXO-15-006}.

For instance, in the right-handed sector, the \Wp\ boson cannot decay into a charged lepton and a hypothetical right-handed 
neutrino if the latter has a mass greater than the \Wp\ boson mass (mixing between \Wp\ and SM $W$ bosons is usually constrained 
to be small from experimental data~\cite{Patrignani:2016xqp}).
 Also, in several theories beyond the SM the \Wp\ boson is expected to couple more strongly to the 
third generation of quarks than to the first and second generations~\cite{Muller:1996dj,StrongTop}.
Searches for a \Wp\ boson decaying into the $t\bar{b}$ final state\footnote{
The notation ``$t\bar{b}$'' is used to describe both the $W'^{+} \to t\bar{b}$ and  $W'^{-} \to \bar{t}b$ processes.}
have been performed at the Tevatron~\cite{Abazov2011145,PhysRevLett.103.041801} in the leptonic top-quark decay channel
and at the Large Hadron Collider (LHC) in both the leptonic~\cite{ATLASWprime7TeV,CMSWprime7TeV,CMSWprime8TeV,ATLASWprime8TeV,Sirunyan:2017vkm}
and fully hadronic~\cite{EXOT-2013-14,Aaboud:2018juj} final states, and the most recent results exclude right-handed \Wp\ bosons with masses
up to about 3.6~\tev\ at the 95\% confidence level.
A previous ATLAS search in the leptonic channel~\cite{ATLASWprime8TeV} using proton--proton ($pp$) collisions at a centre-of-mass energy of $\sqrt{s} = 8$~\tev\ yielded a lower limit of 1.92~\tev\ on the mass of \Wp\ boson with right-handed couplings.
More recently, the CMS Collaboration reported results using a 13~\tev\ $pp$ data set of 35.9\ifb~\cite{Sirunyan:2017vkm}, yielding a lower limit of 3.6~\tev\ on the mass of right-handed \Wp\ bosons.
A search by the ATLAS Collaboration in the fully hadronic decay of the $t\bar{b}$ final state using 36.1\ifb\ of 13~\tev\ data yielded lower limits on the mass of right-handed 
\Wp\ bosons at 3.0~\tev~\cite{Aaboud:2018juj}.  
In each of these analyses, the coupling strength of the \Wp\ boson to right-handed particles was assumed to be equal to the SM weak coupling constant.

\begin{figure}[!h]
  \centering
  \includegraphics[width=0.495\columnwidth]{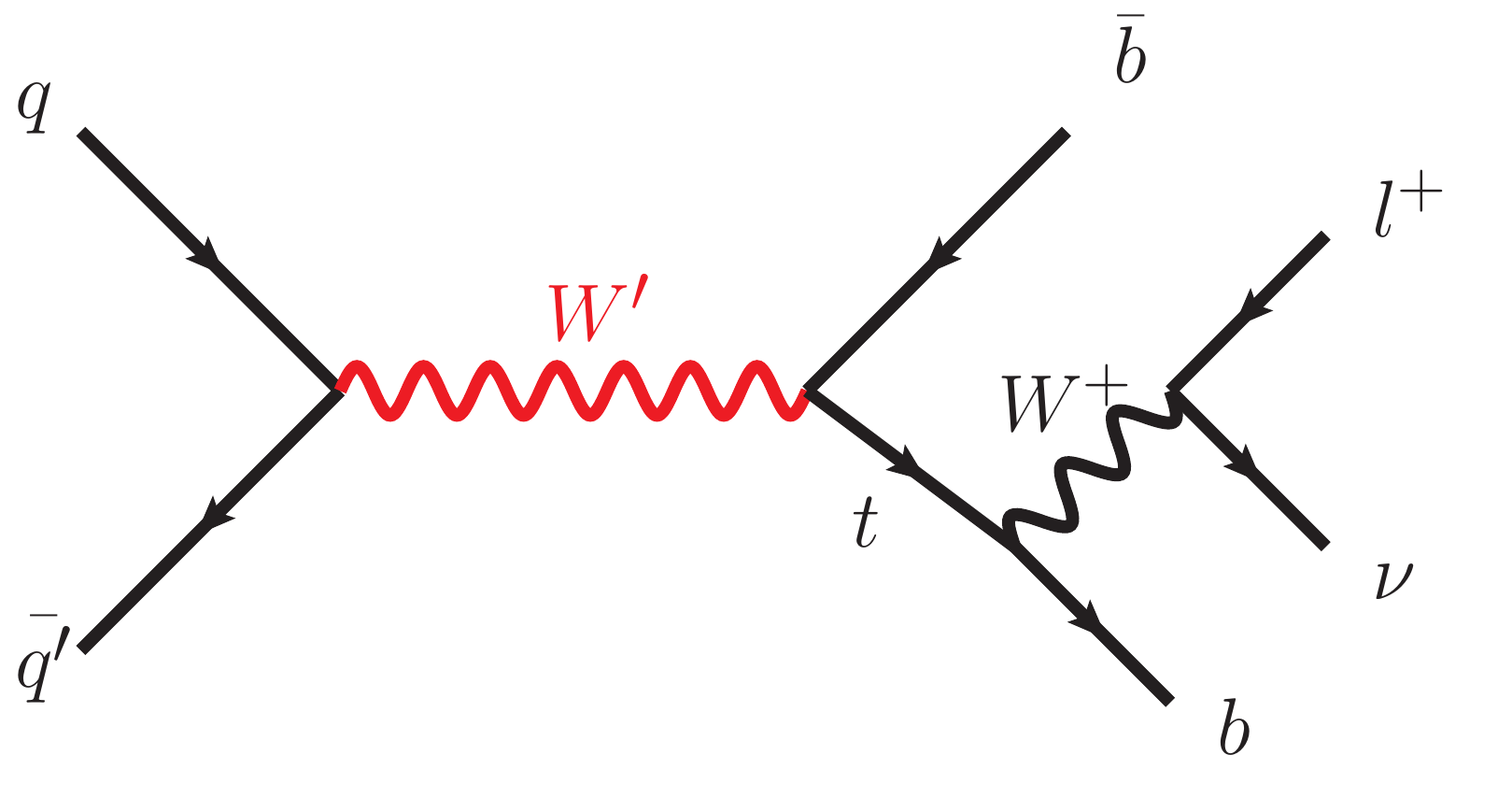}
  \caption{ Feynman diagram for \Wp\ boson production from quark--antiquark annihilation with the subsequent decay into $t \bar{b}$ and a leptonically decaying top quark.}
  \label{fig:wprimeFeyn}
\end{figure}

%-------------------------------------------------------------------------------

This Letter presents a search for \Wp\ bosons using data collected during the period 2015--2016 by the ATLAS detector~\cite{ATLASPaper} at the LHC, corresponding to an integrated luminosity of \lumi\ from $pp$ collisions at a centre-of-mass energy of 13~\TeV.  The search is performed in the $\WpR\ \rightarrow t\bar{b} \rightarrow \ell \nu b\bar{b}$ decay channel, where the lepton, $\ell$, is either an electron or a muon. 
Right-handed \Wp\ bosons, denoted \WpR\ , are searched for in the mass range of 0.5 to 5.0~\tev. A general Lorentz-invariant Lagrangian is used to describe the couplings of the \WpR\ boson to fermions as a function of its mass~\cite{ZackSullivan,Duffty:2012rf}. The mass of the right-handed neutrino is supposed to be larger than the mass of the \WpR\ boson, thus non-hadronic decays of the \WpR\ boson have a negligible branching fraction. In this weakly coupled model, the resulting branching fraction of the \WpR\ to the $t\bar{b}$ final state increases as a function of mass from 29.9\% at 0.5~\TeV\ to 33.3\% at 5~\TeV.

%-------------------------------------------------------------------------------
\section{ATLAS detector}
\label{sec:detector}

The ATLAS detector at the LHC covers almost the entire solid angle around the collision point.\footnote{
ATLAS uses a right-handed coordinate system with its origin at the nominal interaction point in the centre of the detector 
and the $z$-axis along the beam pipe.  
The $x$-axis points from the interaction point to the centre of the LHC ring, and the $y$-axis points upward.  
Cylindrical coordinates ($r, \phi$) are used in the transverse plane, $\phi$ being the azimuthal angle around the beam pipe.  
The pseudorapidity is defined in terms of the polar angle $\theta$ as  $\eta$ = --ln tan($\theta/2$).
Observables labelled ``transverse'' are projected into the $x$--$y$ plane and angular distance is measured in units of
$\Delta R =\sqrt{ (\Delta \eta)^2 + (\Delta \phi)^2}$.} Charged particles in the pseudorapidity range $|\eta| < 2.5$ are reconstructed with the inner detector (ID),
 which consists of several layers of semiconductor detectors (pixel and strip) and a straw-tube transition-radiation tracker, the latter extending to $|\eta| = 2.0$. The high-granularity silicon pixel detector provides
 four measurements per track; the closest layer to the interaction point is known as the insertable B-layer~\cite{IBL1,IBL2}, which was added in 2014 and provides high-resolution hits at small radius to improve the
 tracking performance. The ID is immersed in a 2 T magnetic field provided by a superconducting solenoid.
 The solenoid is surrounded by electromagnetic and hadronic calorimeters, and a muon spectrometer
 incorporating three large superconducting air-core toroid magnet systems. The calorimeter system covers the
 pseudorapidity range $|\eta| < 4.9$. Electromagnetic calorimetry is provided by barrel and endcap high-granularity lead/liquid-argon (LAr) electromagnetic calorimeters, within the region $|\eta|< 3.2$. There is
 an additional thin LAr presampler covering $|\eta| < 1.8$ to correct for energy loss in material upstream of
 the calorimeters. For $|\eta| < 2.5$, the LAr calorimeters are divided into three layers in depth. Hadronic
 calorimetry is provided by a steel/scintillator-tile calorimeter, segmented into three barrel structures within
 $|\eta| < 1.7$, and two copper/LAr hadronic endcap calorimeters, which cover the region $1.5 < |\eta| < 3.2$. The
 forward solid angle out to $|\eta| = 4.9$ is covered by copper/LAr and tungsten/LAr calorimeter modules, which
 are optimized for electromagnetic and hadronic measurements, respectively. The muon spectrometer
 comprises separate trigger and high-precision tracking chambers that measure the deflection of muons
 in a magnetic field generated by the three toroid magnet systems. The ATLAS detector selects events
 using a tiered trigger system~\cite{Aaboud2017:trig}. The first level is implemented in custom electronics and reduces the event
 rate from the LHC crossing frequency of 40 MHz to a design value of 100 kHz. The second level is
 implemented in software running on a commodity PC farm which processes the events and reduces the
 rate of recorded events to 1.0 kHz.

%-------------------------------------------------------------------------------
\section{Data and simulated samples}
\label{sec:samples}

This analysis uses \lumierr\ of $pp$ collisions data at $\sqrt{s} = 13$~\TeV\ recorded using single-electron and single-muon triggers.
Additional data-quality requirements are also imposed, and these are detailed in Section~\ref{sec:objects}. 
During 2015 this corresponded to 3.2\ifb\ with an average of 13.4 interactions per bunch crossing. The 2016 data-taking period corresponds to 32.9\ifb\ with an average of 25.1 interactions per bunch crossing. 

The \WpR\ boson search is performed in the semileptonic decay channel, where the \WpR\ decays into a top quark and a $b$-quark,  the top quark decays into a $W$ boson and a $b$-quark, and the $W$ boson decays in turn into a lepton and a neutrino. The final-state signature therefore consists 
of two $b$-quarks, one charged lepton\footnote{The analysis selects electrons or muons, while the simulation includes $\tau$-leptons. Thus the event yield includes a small contribution due to leptonic decays of $\tau$-leptons.} 
and a neutrino, which is undetected and results in missing transverse momentum, \MET.  The dominant background processes for this signature are therefore the production of $W/Z+$jets (jets arising from light and heavy partons), electroweak single top quarks ($t$-channel, $Wt$ and $s$-channel), $t\bar{t}$ pairs and dibosons ($WW$, $WZ$, and $ZZ$).
An instrumental background due to multijet production, where a hadronic jet is misidentified as a lepton, is also present.
Monte Carlo (MC) simulated events are used to model the \WpR\ signal and all the SM background processes, with the exception of the multijet background prediction, which is derived using data.  The MC generator programs and configurations are summarized in Table~\ref{tab:mcGen}, and described in greater detail in the text below.

Simulated signal events were generated at leading order (LO) by {\textsc{MadGraph5\_aMC@NLO}} v2.2.3~\cite{MadGraph5, Alwall:2014hca, Degrande:2011ua, Alloul:2013bka} using a chiral \WpR\ boson model in which the couplings to the right-handed fermions are like those in the SM. 
{\textsc{MadGraph5\_aMC@NLO}} is also used to model the decays of the top quark, taking spin correlations into account. {\textsc{Pythia8}} v8.186~\cite{Pythia8} was used for parton showering and hadronization, wherein the {\textsc{NNPDF23LO}}~\cite{NNPDF}  parton distribution functions (PDF) of the proton and a set of tuned parameters called the A14 {\textsc{Pythia}} tune~\cite{ATL-PHYS-PUB-2014-021} were used. All samples of simulated events were rescaled to next-to-leading-order (NLO) calculations using NLO/LO $K$-factors ranging from 1.1 to 1.4, decreasing as a function of the mass of the \WpR\ boson, calculated with {\textsc{ZTOP}}~\cite{Duffty:2012rf}. Signal samples were generated between 0.5 and 3~\TeV\ in steps of 250~\GeV, and between 3 and 5~\TeV\ in steps of 500~\GeV.

The benchmark signal model used for this work nominally assumes that the \WpR\ boson coupling strength to fermions, $g^\prime$, is the same as for the $W$ boson: $g^\prime = g$,  where $g$ is the SM SU(2)$_{\textrm{L}}$ coupling.  The coupling of left chiral fermions to the \WpR\ is assumed to be zero.  The total width of the \WpR\ boson increases from 2 to 130~\gev\ for masses between 0.5 and 5~\tev~\cite{ZackSullivan} for $g^\prime = g$ and scales as the square of the ratio $g^\prime/g$. In order to explore the allowed range of the \WpR\ coupling $g^\prime$, samples were also generated for values of $g^\prime/g$ up to 5.0, for several \WpR\ boson mass hypotheses, allowing the effect of increased \WpR\ width to also be included.

Simulated top-quark pair and single-top-quark processes ($t$-channel, $s$-channel and $Wt$) were produced using the NLO {\textsc{Powheg-Box}}~\cite{powheg1,powheg2} generator with the CT10 PDF~\cite{CT10}. The parton shower and the underlying event were added using {\textsc{Pythia}} v6.42~\cite{pythia6} with the Perugia 2012 tune~\cite{PerugiaTunes}. The top-quark pair production MC sample is normalized to an inclusive cross section of $\sigma_{\ttbar}=832 ^{+46}_{-51}$~pb for a top-quark mass of $172.5~\gev$ as obtained from next-to-NLO (NNLO) plus next-to-next-to-leading-logarithm (NNLL) QCD calculations with the Top++2.0 program~\cite{Cacciari:2011hy, Beneke:2011mq, Baernreuther:2012ws, Czakon:2013goa, Czakon:2012pz, Czakon:2012zr, Czakon:2011xx}. 

The background contributions from $W$ and $Z$ boson production in association with jets were simulated
using the {\textsc{Sherpa} v2.2.1}~\cite{sherpa} generator. Matrix elements were calculated for up to two partons at NLO and four partons at LO and merged with the {\textsc{Sherpa}} parton shower using the ME+PS@NLO prescription~\cite{Hoeche:2009rj,Gleisberg:2008fv,Schumann:2007mg}.
The $W/Z$+jets samples are normalized to the inclusive NNLO cross sections calculated with FEWZ~\cite{Hamberg,FEWZ}.

The production of vector-boson pairs ($WW$, $WZ$ or $ZZ$) with at least one charged lepton in the final state was simulated by the {\textsc{Powheg-Box}} generator in combination with {\textsc{Pythia8}} and the leading-order {\textsc{CTEQ6L1}} PDF~\cite{cteq6l}.
The non-perturbative effects were modelled with the {\textsc{AZNLO}} set of tuned parameters~\cite{AZNLO:2014}.

For all {\textsc{MadGraph}} and {\textsc{Powheg}} samples, the {\textsc{EvtGen}}~v1.2.0 program~\cite{EvtGen} was used for the bottom and charm hadron decays.

All simulated event samples include the effect of multiple $pp$ interactions in the same and neighbouring
bunch crossings (pile-up) by overlaying, on each simulated signal or background event,
simulated minimum-bias events generated using {\textsc{Pythia8}}, the {\textsc{A2}} set of tuned parameters~\cite{ATL-PHYS-PUB-2012-003}
and the {\textsc{MSTW2008LO}} PDF set~\cite{Martin:2009iq}.

Simulated samples were processed through the {\textsc{Geant4}}-based ATLAS detector simulation
or through a faster simulation making use of parameterized showers in the calorimeters~\cite{ATLASsim, SAMPLES-G4}.
Simulated events were then processed using the same reconstruction algorithms and analysis chain as used for data.

\begin{table}[!h!tpb]
  \begin{center}\small
   \caption{\label{tab:mcGen} Event generators used for the simulation of the signal and background processes. The PS/Had column describes the program used for parton shower and hadronization.
   }

    \begin{tabular}{llccc}
             \hline \hline
              Process  & Generator & PS/Had & MC Tune & PDF \\
              \hline              
\WpR\ &\textsc{MadGraph5\_aMC@NLO} & \textsc{Pythia8} & A14 & NNPDF23LO\\
$t\bar{t}$ &  \textsc{Powheg-Box} &  \textsc{Pythia6}  & Perugia 2012& NLO CT10 \\
Single-top $t$-channel   & \textsc{Powheg-Box} & \textsc{Pythia6} & Perugia 2012& NLO CT10 \\
Single-top $W+t$         & \textsc{Powheg-Box} &\textsc{Pythia6} & Perugia 2012& NLO CT10 \\
Single-top $s$-channel   & \textsc{Powheg-Box} & \textsc{Pythia6} & Perugia 2012& NLO CT10 \\
$W, Z$    + jets  & \textsc{Sherpa 2.2.1} & \textsc{Sherpa 2.2.1}& Default &NLO CT10 \\
$WW$, $WZ$, $ZZ$  & \textsc{Powheg-Box} &\textsc{Pythia8} & AZNLO & LO CTEQ6L1 \\
             \hline\hline
           \end{tabular}
         \end{center}
 \end{table}

%-------------------------------------------------------------------------------

\section{Event selection and background estimation}
\label{sec:objects}

This search makes use of the reconstruction of multi-particle vertices, the identification and the kinematic properties 
of reconstructed electrons, muons, jets, and the determination of missing transverse momentum. 

Collision vertices are reconstructed from at least two ID tracks with transverse momentum $\pt > 400$~\MeV. 
The primary vertex is selected as the one with the highest $\sum\pt^{2}$, calculated considering all associated tracks.

Electrons are reconstructed from ID tracks that are matched to noise-suppressed topological clusters of energy depositions~\cite{ATL-LARG-PUB-2008-002} in the electromagnetic calorimeter. The clusters are reconstructed using the standard ATLAS sliding-window algorithm, which clusters calorimeter cells within fixed-size rectangles~\cite{PERF-2013-05}. Electron candidates are required to satisfy criteria for the electromagnetic shower shape,  track quality, and track--cluster matching; these criteria are applied using a likelihood-based approach.  Electron candidates must meet the ``Tight'' working point requirements defined in Ref.~\cite{PERF-2013-03} and are further required to have \pt~$>$~25~\GeV\ and  a pseudorapidity of the calorimeter cluster position, $|\eta_{\textrm{cluster}}|$, smaller than 2.47.  Events with electrons falling in the calorimeter barrel--endcap transition region, 1.37~$<$~$|\eta_{\textrm{cluster}}|$~$<$~1.52, which has limited instrumentation, are rejected. 

Muons are identified by matching tracks found in the ID to either full tracks or track segments reconstructed in the muon spectrometer (``combined muons''),  or by stand-alone tracks in the muon spectrometer~\cite{Aad:2016jkr}. They are required to pass identification requirements based on quality criteria applied to the ID and muon spectrometer tracks. Muon candidates must meet the ``Medium'' identification working point requirements defined in Ref.~\cite{Aad:2016jkr}, have a transverse momentum \pt~$>25$~\GeV, and satisfy $|\eta| <2.5$. 

Electron and muon candidates must further satisfy additional isolation criteria that improve rejection of candidates arising from sources other than prompt $W/Z$ boson decays (e.g.\ hadrons mimicking an electron signature, heavy-flavour hadron decays or photon conversions).
Muons are required to be isolated using the requirement that the scalar sum of the \pt\ of the tracks in a variable-size cone around the muon direction
(excluding the track identified as the muon) be less than 6\% of the transverse momentum of the muon. The track isolation cone size is given by
the minimum of $\Delta R = 10 ~\GeV/\pT^\mu$ and $\Delta R$ = 0.3.
Electrons are also required to be isolated using the same track-based variable as for muons, except that the maximum $\Delta R$ in this case is 0.2.
For the purpose of multijet background estimation (see Section~\ref{sec:bgestimation}) electrons and muons satisfying a looser set of identification criteria,
in particular without an isolation requirement, are also considered.

Jets are reconstructed from topological calorimeter clusters using the anti-$k_{t}$ algorithm~\cite{Cacciari:2008gp} 
with a radius parameter of $R = 0.4$, and must satisfy \pT~$>$~25~\GeV\ and  $|\eta|$~$<$~2.5.
To suppress jets originating from in-time pile-up interactions, jets in the range $\pT<60$~\GeV\ and $|\eta|<2.4$ are required to pass the jet vertex tagger~\cite{Aad:2015ina} selection, which has an efficiency of about 90\% for jets originating from the primary vertex.
The closest jets overlapping with selected electron candidates within a cone of size $\Delta R$ equal to 0.2 are
removed from events, as the jet and the electron very likely correspond to the same reconstructed object.
If a remaining jet with \pT~$>$~25~\GeV\ is found close to an electron within a cone of size $\Delta R$~=~0.4, then the electron candidate is discarded. Selected muon candidates near jets that satisfy  $\Delta R$(muon, jet) $< 0.04 + 10~\GeV / \pT^\mu$ are rejected if the jet has at least three tracks 
originating from the primary vertex. Any jets with less than three tracks that overlap with a muon are rejected. 

The identification of jets originating from the hadronization of $b$-quarks (``$b$-tagging'') is based on properties specific to $b$-hadrons, such as long lifetime and large mass.  Such jets are identified using the multivariate MV2c10 $b$-tagging algorithm~\cite{Aad:2015ydr,ATL-PHYS-PUB-2016-012},  which  makes use of information about the jet kinematic properties, the characteristics of tracks within jets, and the presence of displaced secondary vertices. The algorithm is used at the $77\%$ efficiency working point and provides a rejection factor of 134 (6.21) for jets originating from light-quarks or gluons (charm quarks), as determined in simulated \ttbar\ events. 
Jets satisfying these criteria are referred to as ``$b$-tagged'' jets.

The presence of neutrinos can be inferred from an apparent momentum imbalance in the transverse plane.
The missing transverse momentum (\met) is calculated as the modulus of the negative vectorial sum
of the transverse momentum of all reconstructed objects (electrons, muons, jets) as well as
specific ``soft terms'' considering tracks associated with the primary vertex that do not match the selected reconstructed objects~\cite{Aaboud:2018tkc}.

Candidate events are required to have exactly one charged lepton, two to four jets with at least one of them $b$-tagged and a minimum \MET\ threshold that depends on the lepton flavor.  From these objects, $W$ boson and top-quark candidates are reconstructed and final requirements on the event kinematic properties are applied to define several orthogonal regions with enriched signal content, as well as signal-depleted regions to validate data modelling.  
The jet, $b$-tag and lepton requirements define basic selections, which are labelled as $X$-jet $Y$-tag where $X=2, 3, 4$ and $Y=1, 2$, separated for electron and muon channel selections.   

The $W$ boson candidate is reconstructed from the lepton and \MET, with the assumption that only one neutrino is present in the event.
The $z$ component of the neutrino momentum ($p_{z}$) is calculated from the invariant mass of the lepton--\MET\ system with the constraint that $m_W=80.4$~\GeV. The constraint yields a quadratic equation and in the case of two real solutions, the smallest $|p_{z}|$ solution is chosen.  If the transverse mass, $m_{\textrm{T}}^W$, of the reconstructed $W$ boson is larger than the value $m_W$ used in the constraint, the two solutions are imaginary. This case can be due to the resolution of the missing transverse momentum measurement.  Here, the $E_{x,y} ^{\text{miss}}$ components are adjusted to satisfy $m_{\textrm{T}}^W = m_W$, yielding a single real solution.

The four-momentum of the top-quark candidate is reconstructed by adding the four-momenta of the $W$-boson candidate and of the jet, 
among all selected jets in the event, that yields the invariant mass closest to the top-quark mass ($m_{\textrm{top}}=172.5~\GeV$). 
Thereafter, this jet is referred to as ``$b_{\text{top}}$'', and may not be the jet actually $b$-tagged. Finally, the four-momentum of the candidate \Wp\ boson is reconstructed by adding the four-momentum of the reconstructed top-quark candidate and the four-momentum of the highest-\pt\ remaining jet (referred to as ``$b_1$'').  The \Wp\ four-momentum is used to evaluate the invariant mass of the reconstructed $W^\prime \rightarrow tb$ system ($m_{tb}$), which is the variable used for background discrimination for this search.

An event selection common to all signal and validation regions is defined as: lepton $\pt>50$~\GeV, $\pt(b_1)> 200$~\GeV, $\pt({\textrm{top}})> 200$~\GeV, and $\MET>30$~\GeV.
In order to keep the multijet background at a low level an additional selection is imposed, in the muon channel, on the sum of $m_{\textrm{T}}^W$ and \MET: $m_{\textrm{T}}^W + \MET>100~\GeV$.
In the electron channel the same requirement is applied to keep the selection in both channels as similar as possible, and, in addition the $\MET$ threshold is raised to $80$~\GeV\ to further suppress the multijet background.
This phase space is then subdivided into a signal region (SR), a validation region enriched with the $W$+jets background (VR$_{\textrm{pretag}}$), a validation region enriched with the \ttbar\ background (VR$_{\ttbar}$), and a validation region enriched with the $W$+heavy-flavour jets background (VR$_{\textrm{HF}}$).
All regions consist of events with two or three jets, except for the VR$_{\ttbar}$ where events with exactly four jets are selected.
The SR and VR$_{\ttbar}$ require that one or two jets are $b$-tagged, while only one $b$-tagged jet is required in the VR$_{\textrm{HF}}$.
No $b$-tagging requirement is applied in the VR$_{\textrm{pretag}}$.
Specific selections are then applied in the two following cases.
The SR is defined by requiring that the angular separation of the lepton and $b_{\text{top}}$ be small: $\Delta R(\ell, b_{\text{top}})<1.0$. An additional criterion $m_{tb}>500$~\GeV\ is applied to remove a small number of low-mass $W$+jets and \ttbar\ events. 
The VR$_{\textrm{HF}}$ consists of events where the lepton--jet and jet--jet separations are large: $\Delta R(\ell, b_{\text{top}})>2.0$ and $\Delta R(b_1, b_{\text{top}})>1.5$.  
The application of these two selections reduces the $t\bar{t}$ background in the VR$_{\textrm{HF}}$ region by 90\%.
The expected signal contamination in the validation regions is at most 5\% for low \Wp\ masses, and falls below $10^{-4}$ for \Wp\ masses above 3~\TeV.
The event selection criteria for each region are summarized in Table~\ref{tab:evtselection}.

\begin{table}[h]
\caption{\label{tab:evtselection} 
Summary of the event selection criteria used to define signal and validation regions.  The \MET selection cut is harder for events with electrons than with muons.
}
\begin{center}
\begin{tabular}{c||c|c|c}
\hline
\multicolumn{4}{c}{Common selection} \\
\hline
\multicolumn{4}{c}{ \pt $(\ell)>$~50~\GeV, \pt $(b_1)> 200$~\GeV, \pt $({\textrm{top}})> 200$~\GeV} \\
\multicolumn{4}{c}{\MET $>$30 (80)~\GeV, $m_{\textrm{T}}^W + \MET>100~\GeV$}\\
\hline
\hline
Signal Region & VR$_{\textrm{pretag}}$ & VR$_{\ttbar}$ & VR$_{\textrm{HF}}$ \\
\hline
2 or 3 jets                      & 2 or 3 jets   & 4 jets          & 2 or 3 jets \\
1 or 2 $b$-jets                  & pretag        & 1 or 2 $b$-jets & 1 $b$-jet \\
$\Delta R(\ell,b_{\text{top}})<1.0$ &               &                 & $\Delta R(\ell, b_{\text{top}})>2.0$ \\ 
$m_{tb}>500$~\GeV                 &               &                 & $\Delta R(b_1, b_{\text{top}})>1.5$\\ 
\hline
\end{tabular}
\end{center}
\end{table}

The signal selection efficiency (defined as the number of events passing all selection requirements divided
by the total number of simulated $W' \to t\bar{b} \to \ell \nu b \bar{b}$ events) in the signal region
is shown, as a function of the simulated \WpR\ mass, in Figure~\ref{fig:SignalAcceptanceWpR}. 
Selection efficiency curves are shown for the electron and muon channels separately, as well as for the pretag selection. 
Due to the jet $p_{\textrm T}$ and $\Delta R(\ell, b_{\textrm{top}})$ requirements, the signal has vanishing efficiency for a \WpR\ mass of 500~\gev, 
but the efficiency rises as the decay products become more boosted.
The maximum SR signal efficiency, 11.3\%, is obtained for a mass of 1.5~\tev, then the efficiency decreases for higher masses to 5.3\% at 5~\tev. 
The application of the $b$-tagging requirement has a larger impact on the signal efficiency at high \WpR\ boson mass values.
In the electron channel the electron--jet overlap criterion does not allow
the electron to be close ($\Delta R(\ell,{\textrm{jet}})<0.4$) to the jet. In the muon channel, this criterion is relaxed
by using a variable $\Delta R$ cone size, resulting in an improved signal acceptance. 

\begin{figure}[!h]
  \centering
     \includegraphics[width=0.70\columnwidth]{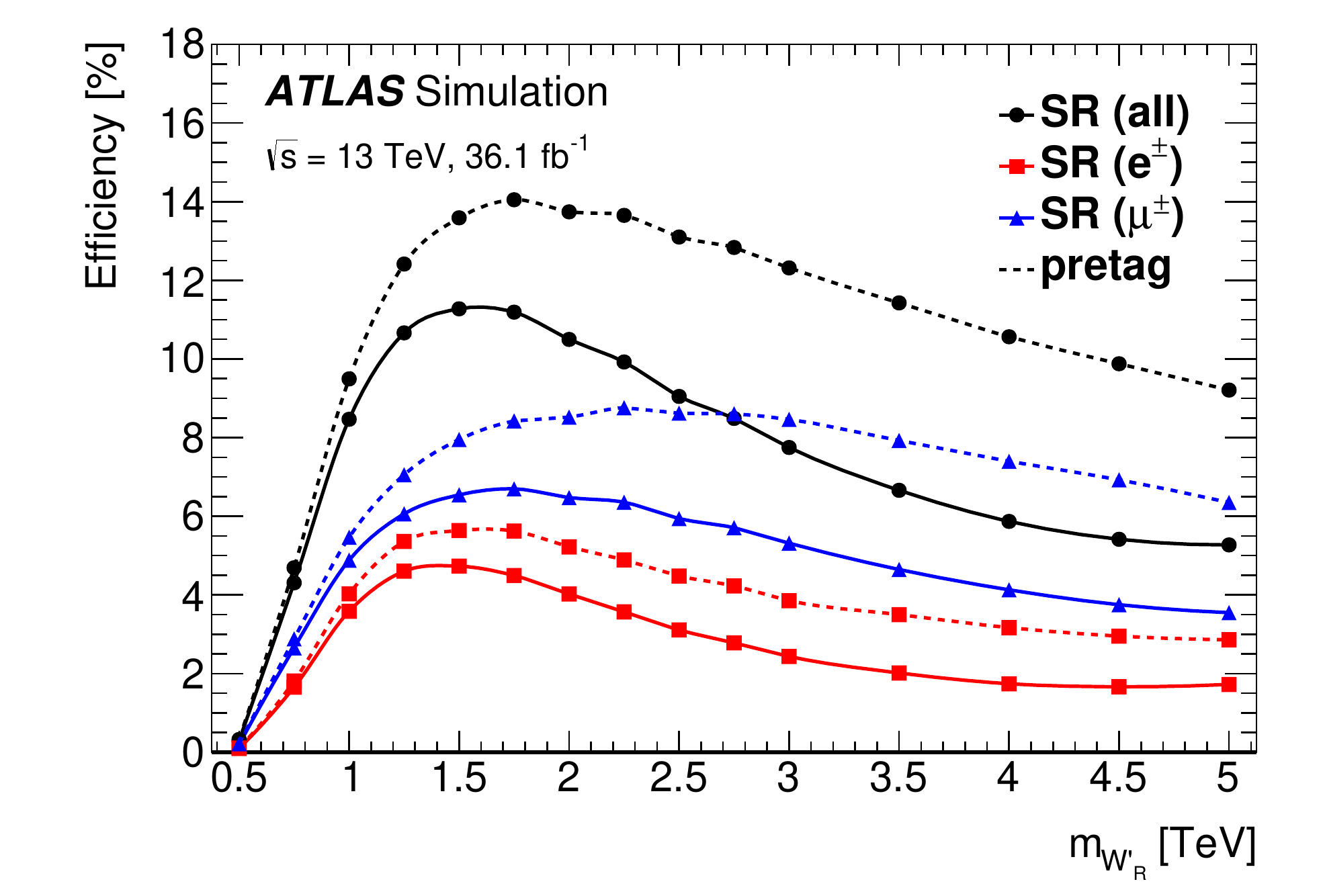}
     \caption{Signal selection efficiency (efficiency is defined as the number of events passing all selections divided by the total number of simulated $W' \to t\bar{b} \to \ell \nu b \bar{b}$ events) in the signal region as a function of the simulated \WpR\ mass. 
Efficiencies are shown for: all channels combined (full circle), electron channels only (full square) and muon channels only (full triangle).
For reference, signal efficiency curves are also shown without the requirement on $b$-tagging (pretag selection: dotted lines).}
  \label{fig:SignalAcceptanceWpR}
\end{figure}

\section{Background estimation}
\label{sec:bgestimation}

The \ttbar,  single-top-quark, diboson and $W/Z$+jets backgrounds are modelled using the simulated MC samples and are normalized to the theory predictions of the inclusive cross sections, while the multijet background is estimated using the data as described below in this section.  Each of these background samples gives rise to individual differential $m_{tb}$ templates predicting their unique kinematic properties.  These initial background normalizations are taken as starting values, and the final normalization is determined through a maximum-likelihood fit of the background templates to the data in which the background normalizations are parameters of the fit (described in Section~\ref{sec:result}).  Because the signal regions are dominated by \ttbar\ and $W$+jets production, the normalization of these backgrounds is allowed to float freely in the maximum-likelihood fit with no prior.

The background arising from multijet production consists of events with a jet that is misreconstructed as a lepton or with a non-prompt lepton that satisfies the lepton identification criteria.  The simulation of this background source is challenging as it suffers from large systematic uncertainties and does not reliably reproduce the observed data in regions enriched with multijet events.  
Therefore the multijet background is estimated from data with the so-called {\em matrix method}, which is used to disentangle the mixture of non-prompt leptons found in the multijet background and prompt leptons originating from $W$/$Z$ bosons~\cite{TOPQ-2011-02}.  
This method uses a data sample, with loosened identification criteria, dominated by multijet production and with a small contamination of electroweak (EW) $W/Z+$jets production.
The probability that a jet from multijet production which passes the loose selection also satisfies the tight selection criteria is estimated in this control region.
The multijet purity in this sample is improved by subtracting, using MC simulation, the EW contamination to remove bias due to prompt-lepton sources.
The efficiency for prompt leptons passing the loose selection to also pass the tight selection is determined using $t\bar{t}$ MC samples, 
corrected using comparisons of MC and data $Z \to \ell\ell$ events.
The number of multijet background events satisfying the selection criteria is estimated from these efficiencies using data events that satisfy all criteria, 
except that loose lepton identification criteria are used.
While this data-driven method is a significant improvement on the use of MC simulation, the low number of events and inherent systematic variations of the EW contribution lead to a significant systematic uncertainty.  
Systematic uncertainties on the multijet background are evaluated~\cite{Aaboud:2018mjh} using various definitions of multijet control regions 
and by considering systematic uncertainties associated with object reconstruction and MC simulation.
The uncertainty on this background is taken as 50\% of the total rate and treated as uncorrelated between selected regions.

Figure~\ref{fig:VRmtb} shows the distributions of the reconstructed invariant mass of the \Wp\ boson candidate for data and for background predictions in the 2-jet 1-tag VR$_{\textrm{HF}}$ and 4-jet 2-tag VR$_{\ttbar}$ validation regions.
Background templates are fit to data in each VR using the same statistical method as for the signal region
except that the normalizations of $t\bar{t}$ and $W$+jets backgrounds
are constrained to the post-fit rates obtained in the signal region (see Section~\ref{sec:result}).

\begin{figure}[hbtp]
\centering
\includegraphics[width=0.40\textwidth,height=0.40\textwidth]{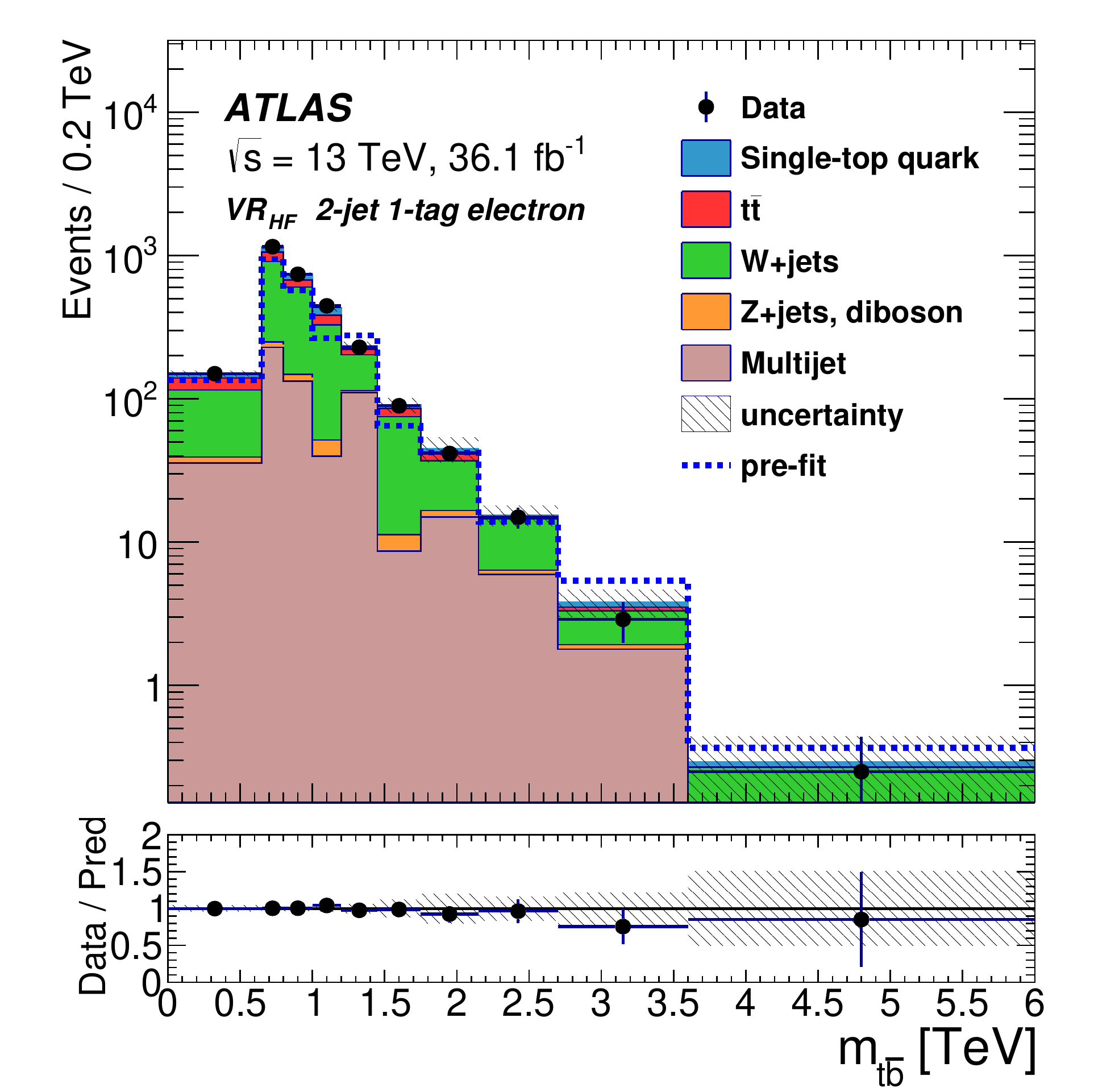}
\includegraphics[width=0.40\textwidth,height=0.40\textwidth]{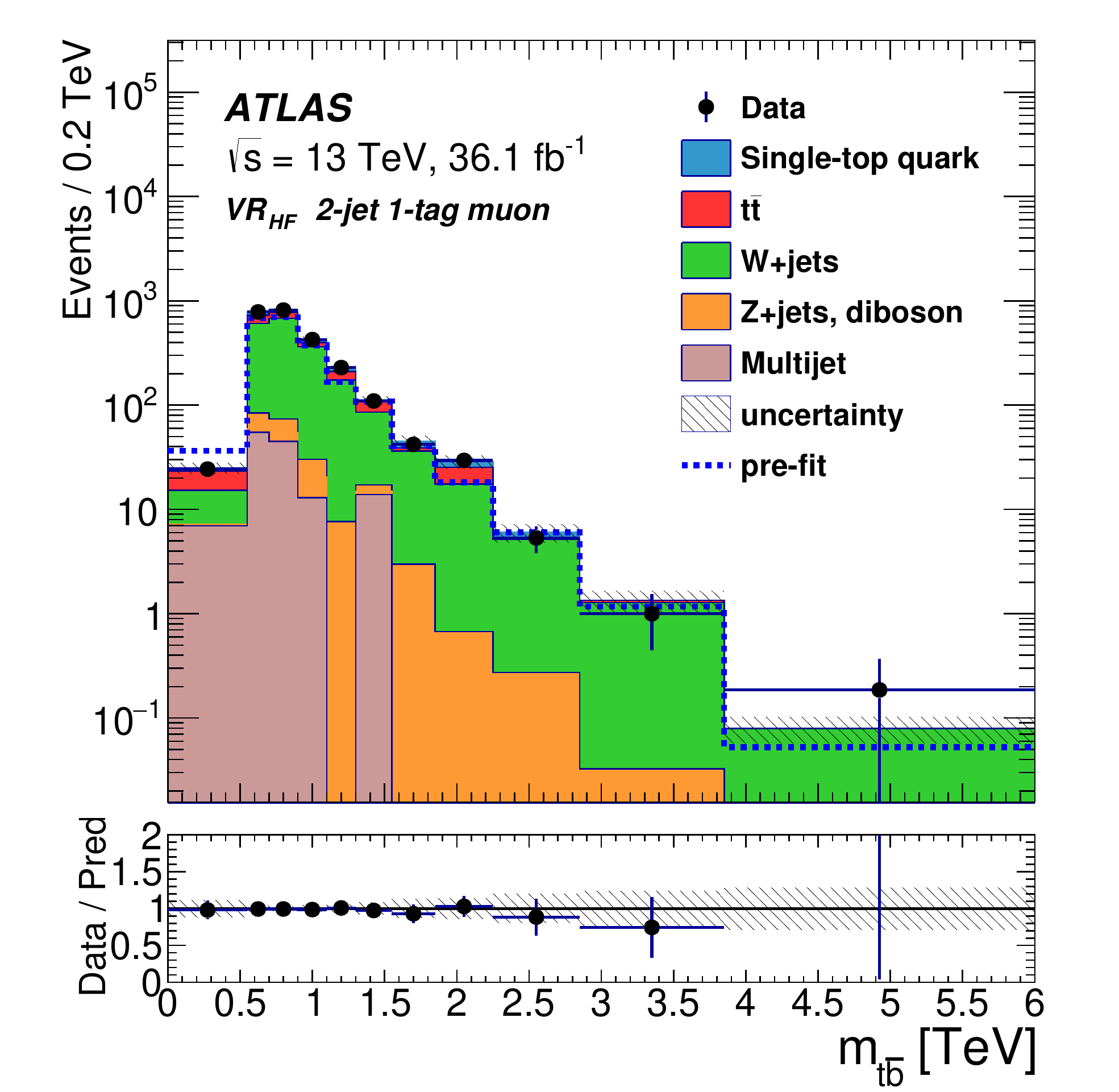}\\
\includegraphics[width=0.40\textwidth,height=0.40\textwidth]{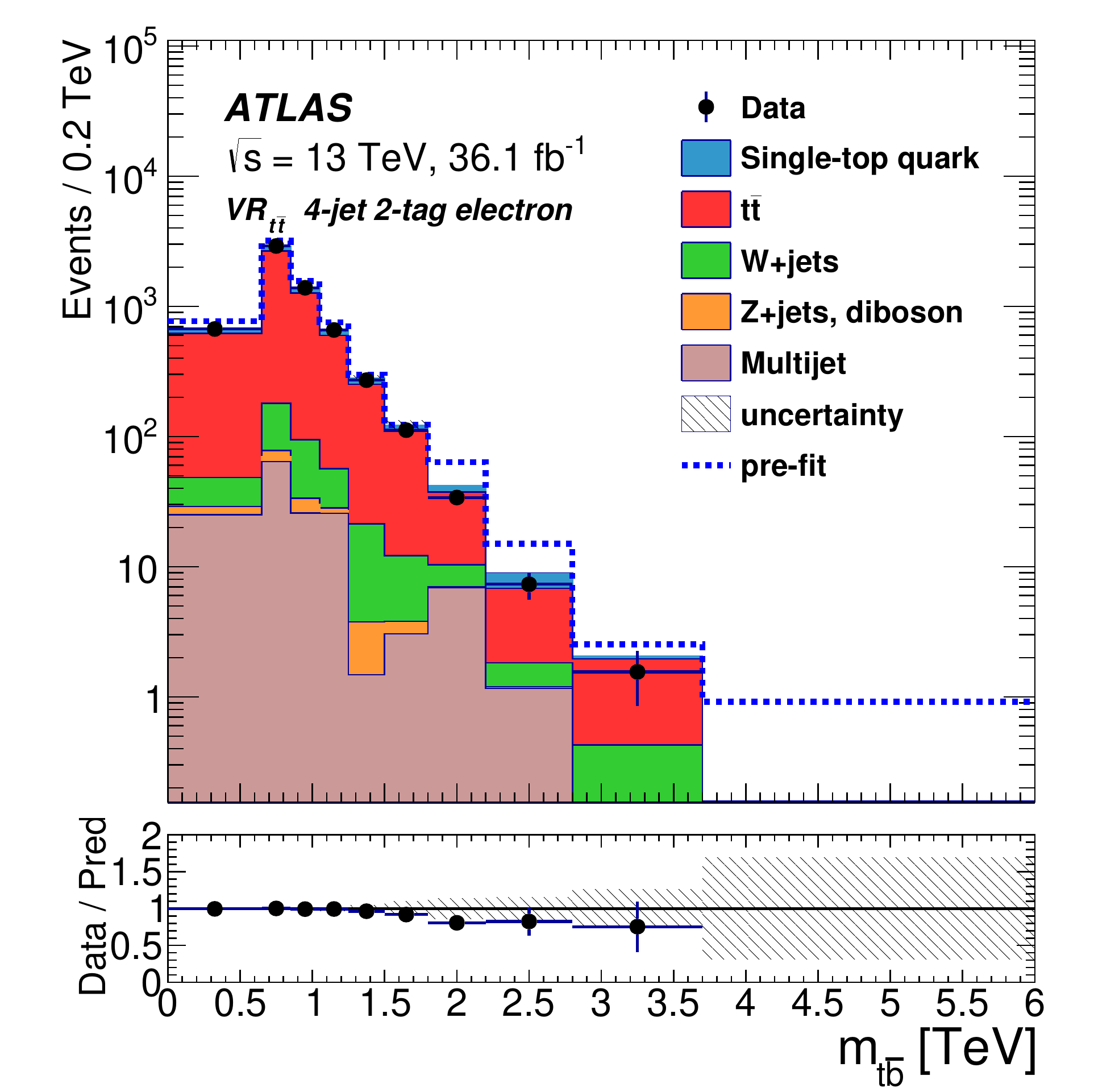}
\includegraphics[width=0.40\textwidth,height=0.40\textwidth]{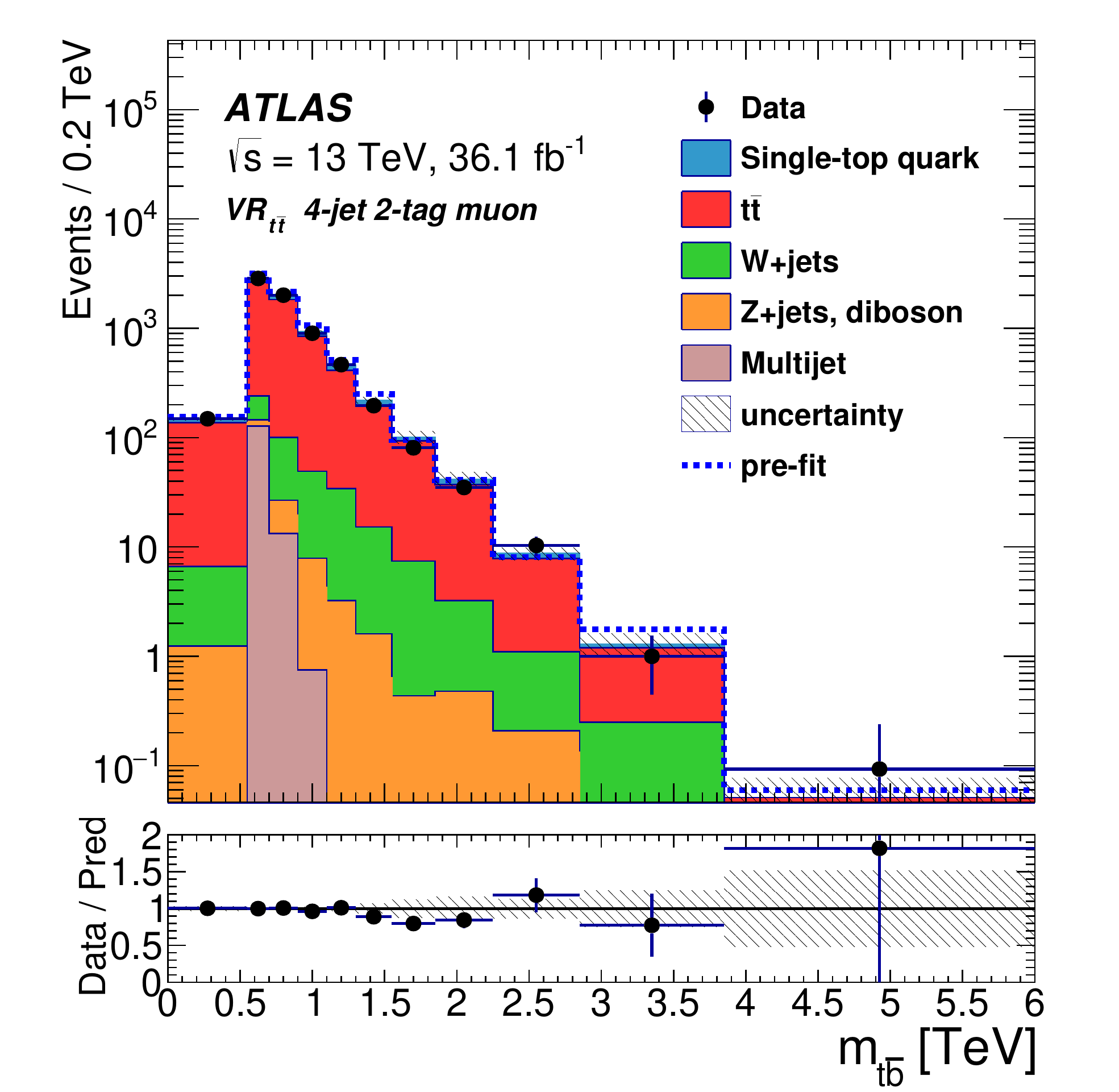}  
\caption{
  Distributions of the reconstructed invariant mass of the \Wp\ boson candidate in the (top) 2-jet 1-tag VR$_{\textrm{HF}}$ and (bottom) 4-jet 2-tag VR$_{\ttbar}$ validation regions.
  Background templates are fit to data in each VR using the same statistical method as for the signal region except that the normalizations of $t\bar{t}$ and $W$+jets backgrounds are constrained to the post-fit rates obtained in the signal region (see Section~\ref{sec:result}). 
The pre-fit line presents the background prediction before the fit is performed. 
Uncertainty bands include all the systematic and statistical uncertainties. The residual difference between the data and MC yields is shown as a ratio in the bottom portion of each figure, wherein the error bars on the data points correspond to the data Poisson uncertainty.
}
\label{fig:VRmtb}
\end{figure}

%-------------------------------------------------------------------------------
\section{Systematic uncertainties}
\label{sec:systematics}

Two primary sources of systematic uncertainty, experimental and modelling, affect the reconstruction of the $m_{tb}$ distributions. 
Experimental uncertainties arise due to the trigger selection, the object reconstruction and identification, as well as the object energy, momentum and mass calibrations and their resolutions. 
Modelling uncertainties result in shape and normalization uncertainties of the different MC samples used to model the signal and backgrounds. These stem from uncertainties in the generator matrix-element calculation, the choice of parton shower and hadronization models and their parameter values, the PDF set and the choice of renormalization and factorization scales.  
The impact on the signal and background event yields of the main systematic uncertainties is summarized in Table~\ref{tab:syst}, wherein the uncertainty on the overall yield is presented for each background source. 
All values are given as a percentage change in overall yield and represent the prior values assigned before fitting. 
The source of each uncertainty is described in this section, and uncertainties are considered fully correlated across all eight signal regions and among processes, unless specifically noted. 

\begin{table}[hbtp]
\caption{\label{tab:syst} 
Impact of the main sources of uncertainty on the signal and background event yields.
All values are given as a percentage change in the overall yield and represent the prior values assigned before fitting. 
Uncertainties for the signal are given for a \WpR\ mass hypothesis of 2~\TeV.
Uncertainties in the background are the same for all signal masses.
Systematic uncertainties in the normalization, 2-jet vs 3-jet region cross-extrapolation, and reconstructed $m_{tb}$ shape of the signal and background processes 
are described in the text.  
Sources of uncertainty may affect both the total event yield and the shape of the $m_{tb}$ distribution. 
An ``S'' indicates that a shape variation has been included, in addition to the rate variation, due to the sources listed.
``U'' refers to regions that are not correlated with one another and  ``F'' refers to a normalization that floats freely. In certain instances of freely floating normalizations, the rate variation of systematic effects is removed, thus leaving only a shape variation.  Such cases are indicated with a "*" symbol.
The ``Jets'' column includes uncertainties related to \MET.
A range of values correspond to the lowest and the highest values determined across different channels in the SR. The final column describes the uncertainty in extrapolating event yields between the 2-jet and 3-jet selections.
}
\begin{center}
 \begin{adjustwidth}{-0.75cm}{}
\begin{tabular}{l c c S[table-format=1] @{--} S[table-format=2,table-number-alignment = left] S[table-format=1] @{--} S[table-format=2,table-number-alignment = left] ccc}
\hline
\hline
Process 			& Norm.	& Lumi. 	&  \multicolumn{2}{c}{$b$-Tagging [S]} 	&  \multicolumn{2}{c}{Jets [S]}		&  Leptons 	& Modelling [S] 		& 2j/3j Extrap. \\
\hline
\WpR\ 			& - 		& 2.1		& \;\;\;\;\;\,8&12  			& \;\;1&4  		& 1--2		& - 				& - \\
\ttbar 			& F		& - 		& \;\;\;\;\;\,2&6 			& \;\;4&8 		& 1--2		& 0 [*]			&- \\
$W+$jets 			& F 		& - 		& \;\;\;\;\;\,6&15 			& \;\;2&12 		& 1--3		& 	0 [*]			&20 \\
$Z+$jets 			& \tablenum[table-format=2]{20}		& 2.1		& \;\;\;\;\;\,6&12 			& \;\;2&9 		& 1--3 		& - 				&- \\
Diboson 			& \tablenum[table-format=2]{11} 		& 2.1 	& \;\;\;\;\;\,3&10  			& \;\;2&8 		&1--2 		& - 				& - \\
Single top quark	& \tablenum[table-format=2]{6} 		& 2.1 	& \;\;\;\;\;\,2&7 			& \;\;1&4 		&1--2		& 6--22 			& - \\
Multijet 			&  \;\;\;\;\;\,50  [U]	& - 		& \alignUnderEnDash{-}	&			&\alignUnderEnDash{-}	& 			&- 			& - 				& - \\
\hline
\hline
\end{tabular}
 \end{adjustwidth}
\end{center}
\end{table}

The selection of jets and \MET\ has an associated uncertainty related to the calorimeter calibration of the energy scale and the calorimeter resolution, as well as to the identification/reconstruction efficiencies of objects reconstructed using the calorimeter, sample flavour composition and corrections for pile-up and neutrinos produced in hadron decays. The uncertainty contributed by each source is typically 1--5\% of the expected event rates and can impact the shape of differential distributions.  In addition, the \MET\ calculation leads to a typical uncertainty in the event yield of less than 1\%.

The process of $b$-tagging jets has an uncertainty in the scale factors required to match the tagging efficiency between data and simulation.  These uncertainties are evaluated independently for jets arising from $b$-quarks, $c$-quarks and light-quarks or gluons.  The uncertainty in the selection efficiency for tagging $b$-quarks is typically small (1--5\% per jet) except for very high \pt\ jets where it can increase to 6\% per jet, and the mis-tagging of $c$-/light-quarks and gluons can be as large as 10\%.  These sources of uncertainty can additionally induce non-uniform variations in differential distributions of up to 10\%.

The uncertainty in the reconstruction efficiency and acceptance of leptons due to trigger, reconstruction and selection efficiencies in simulated samples is roughly 1\% of the total event yield. The energy/momentum scale and resolution for leptons is corrected in simulation to match data measurements, and the resulting uncertainty in the efficiency arising from these corrections is less than 1--2\%.  

The normalization of simulated samples has an associated uncertainty that varies by production process. The uncertainty in the cross section times branching fraction for single-top and diboson production is taken as 6\%~\cite{Kidonakis:2011wy, Kidonakis:2010tc, Kidonakis:2010ux} and 11\%~\cite{DibosonUnc}, respectively. An uncertainty of 20\% is assumed for $Z$+jets rate, which represents a very small background, in line with the modelling uncertainty assigned for $W$+jets (see below in this section).
The cross sections for the \ttbar\ and $W+$jets samples are normalized using freely floating parameters whose values are determined by fitting to data.  All simulated samples that are normalized to the ATLAS luminosity measurement are assigned a luminosity uncertainty of 2.1\%.  
This uncertainty is derived, following a methodology similar to that detailed in Ref.~\cite{DAPR-2013-01}, from a calibration of the luminosity scale using $x$--$y$ beam-separation scans performed in August 2015 and May 2016.

Differences due to the choice of MC generator, fragmentation/hadronization model, and initial/final-state radiation model are treated as a source of uncertainty for the \ttbar\ and $t$-channel single-top-quark simulations.
The uncertainty due to the choice of MC generator is evaluated as the difference in yield between the nominal choice of {\textsc{Powheg-Box}} and the alternative {\textsc{MadGraph5\_aMC@NLO}}~\cite{aMCNLO} generator, using \Herwigpp~\cite{Bahr:2008pv, Bellm:2015jjp} for showering in both instances.
The uncertainty due to the fragmentation/hadronization model is evaluated by comparing {\textsc{Pythia6}} and \Herwigpp\ simulated samples.
Variations of the amount of additional radiation are studied by changing the scale of the hard-scatter process and the scales in parton-shower simulation
simultaneously using the {\textsc{Powheg-Box}}+{\textsc{Pythia6}} set-up. In these samples, a variation of the factorization
and renormalization scales by a factor of two is combined with the Perugia2012radLo tune and a variation
of both scales by a factor of 0.5 is combined with the Perugia2012radHi tune~\cite{PerugiaTunes}. In the case of $t\bar{t}$ production the
{\textsc{Powheg-Box}} $h_{\textrm{damp}}$ parameter, which controls the transverse momentum of the first additional emission beyond the Born configuration,
is also changed simultaneously, using values of $m_{\textrm{top}}$ and $2 \times m_{\textrm{top}}$, respectively.
An uncertainty associated with the NLO calculation of $Wt$ production~\cite{WtProd} is evaluated by comparing the baseline sample generated with the diagram removal scheme
to a $Wt$ sample generated with the diagram subtraction scheme.

These differences yield relative variations in shape and normalization of 1--3\% on average, although the variation can be larger than 10\% in the highest $m_{tb}$ regions probed.  The normalization component of these modelling uncertainties is removed for the \ttbar\ samples because the overall normalization is determined via the data in this case.

Differences between the predictions for the ratio of 2-jet to 3-jet yields from different showering simulations were studied for the \ttbar\ and $W$+jets simulation. These differences are estimated by simultaneously varying the renormalization and factorization scales, and by using different MC generators. While only small differences were observed for \ttbar\ simulation, the ratio of the yields of 2-jet to 3-jet selections in $W$+jets simulation varied by up to 20\%.  Thus, an additional uncertainty of 20\% is assigned to the $W$+jets yield in the 3-jet selection.\footnote{For the $Z$+jets background a similar variation could be expected, but since this background is minor, a 20\% constant rate uncertainty is simply assumed.} 

Uncertainties in $W$+jets modelling are determined by comparing the nominal {\textsc{Sherpa}} simulation with an alternative sample produced with the {\textsc{MadGraph5\_aMC@NLO}} generator interfaced to {\textsc{Pythia8}} for parton showering and hadronization.  The uncertainty in our knowledge of the flavour fraction in the $W$+jets sample is tested by splitting the $W$+jets sample into light-quark/gluon and heavy-flavour components and by decorrelating the $W$+jets shape uncertainty between 2-jet and 3-jet events. In each case, no significant effect on the extracted results is observed.

The uncertainty in the yield of simulated \ttbar\ background events due to the choice of PDF is evaluated using the PDF4LHC recommendations~\cite{PDF4LHC15}.  The statistical uncertainty of the limited MC samples is included in each histogram bin of the $m_{tb}$ distribution.

%-------------------------------------------------------------------------------
\section{Results}
\label{sec:result}

In order to test for the presence of a massive resonance, the $m_{tb}$ templates obtained from the signal and background simulated event samples are fit to data using a binned maximum-likelihood (ML) approach based on the \textsc{RooStats} framework ~\cite{Moneta:1289965,Verkerke:2003ir,Baak:2014wma}.  Each signal region selection is considered simultaneously as an independent search channel, for a total of eight regions corresponding to mutually exclusive categories of electron and muon, 2-jet and 3-jet, and 1-$b$-tag and 2-$b$-tags. 

The normalizations of the \ttbar\ and $W+$jets backgrounds are free parameters in the fit, while other background normalizations are assigned Gaussian priors based on their respective normalization uncertainties.  The systematic uncertainties described in Section~\ref{sec:systematics} are incorporated in the fit as nuisance parameters with correlations across regions and processes taken into account. The signal normalization is a free parameter in the fit.

The expected and observed event yields after the ML fit are shown in Tables~\ref{tab:SR1tag} and~\ref{tab:SR2tag} and correspond to an integrated luminosity of \lumi. 
The fitted \ttbar\ and $W+$jets rates relative to their nominal predictions are found to be $0.98\pm0.04$ and $0.78\pm0.19$, respectively.
For these two backgrounds the total uncertainty reported in the event yield tables is smaller than the uncertainty 
in the fitted normalization factor because there are anticorrelations between nuisance parameters in 
the likelihood fit.

The  $m_{tb}$ distributions for the SR after the ML fit are shown in Figures~\ref{fig:SRmtb2jet} and~\ref{fig:SRmtb3jet}. 
An expected signal contribution corresponding to a \WpR\ boson with a mass of 2.0~\tev\ is shown as a dashed histogram overlay.
The binning of the $m_{tb}$ distribution is chosen to optimize the search sensitivity while minimizing statistical fluctuations. 
Requirements are imposed on the expected number of background events per bin, and the bin width is adapted to a resolution function that represents the 
width of the reconstructed mass peak for each studied \WpR\ boson signal sample.

For a \WpR\ boson with a mass of 2~\TeV\ and nominal $g'/g=1$ coupling the total expected uncertainty in estimating the signal strength\footnote{The signal strength is defined as the ratio of the signal cross section estimated using the data to the predicted signal cross section.} is 12\%. The total systematic uncertainty is 9\%, and the largest uncertainties are due to the $t\bar{t}$ generator (4.0\%), jet energy scale (JES) (2.8\%), $t\bar{t}$ showering (2.5\%), $t\bar{t}$ normalisation (2.0\%) and JES $\eta$ intercalibration modelling (1.3\%).
For resonances with a mass of 2.5~\TeV\ or above, the data Poisson uncertainty becomes the largest uncertainty in estimating the signal rate, while the total systematic uncertainty is dominated by the uncertainty on the $b$-tagging efficiency.

\begin{table}[hbtp]\small
  \setlength{\tabcolsep}{0.4em}
\begin{center}
\caption{\label{tab:SR1tag} 
The numbers of signal and background events and the numbers of observed data events are shown in the 2-jet 1-tag and 3-jet 1-tag  signal regions.
For signal, the values correspond to expected event yields and quoted uncertainties account for the statistical uncertainty of the number of events in the simulated samples. 
The number of background events is obtained following a ML fit to the data and uncertainties contain statistical and systematic uncertainties.
}
\begin{tabular}{l S[table-number-alignment=center] p{0.05cm} S[table-column-width=3em,table-number-alignment=center] p{0.4cm} S[table-column-width=3em,table-number-alignment=center]  p{0.05cm} S[table-column-width=3em,table-number-alignment=center] p{0.4cm} S[table-column-width=3em,table-number-alignment=center] p{0.05cm} S[table-column-width=3em,table-number-alignment=center] p{0.4cm}S[table-column-width=3em,table-number-alignment=center] p{0.05cm} S[table-column-width=3em,table-column-width=3em,table-number-alignment=center] }
\hline\hline
& \multicolumn{3}{c}{2-jet 1-tag ($e^\pm$)} & & \multicolumn{3}{c}{2-jet 1-tag ($\mu^\pm$) }  & & \multicolumn{3}{c}{3-jet 1-tag ($e^\pm$)} & & \multicolumn{3}{c}{3-jet 1-tag ($\mu^\pm$)}\\
\hline
$W'_\text{R}$ (1.0 TeV) &     1517 & $\pm$ &        32 & &     2030 & $\pm$ &        40 & &    1159 & $\pm$ &        31 & &     1665 & $\pm$ &        35\\ 
$W'_\text{R}$ (2.0 TeV) &       83.4 &  $\pm$ &       1.7 & &     132.9 & $\pm$ &        2.1 & &     105.0 & $\pm$ &        1.9 & &     167.4 & $\pm$ &        2.2\\ 
$W'_\text{R}$ (3.0 TeV) &        4.7 & $\pm$ &       0.1 & &      10.4 & $\pm$ &       0.2 & &      7.0 & $\pm$ &       0.2 & &      15.7 & $\pm$ &       0.2\\ 
$W'_\text{R}$ (4.0 TeV) &        0.43 & $\pm$ &       0.01 & &       1.01 & $\pm$ &       0.02 & &        0.64 &  $\pm$ &      0.02 & &        1.62 &  $\pm$ &      0.03\\ 
$W'_\text{R}$ (5.0 TeV) &        0.076 & $\pm$ &      0.002 & &        0.153 &  $\pm$ &      0.003 & &        0.096 & $\pm$ &       0.003 & &        0.232 & $\pm$  &      0.004\\ \hline
             $t\bar{t}$ &  1112  & $\pm$ & 23 &&  1505  & $\pm$ & 28 &&  3220  & $\pm$ & 50 &&  4090  & $\pm$ & 70  \\ 
            Single-top  &   472  & $\pm$ & 20 &&   657  & $\pm$ & 25 &&   482  & $\pm$ & 21  &&  624  & $\pm$ & 24 \\   
               $W$+jets &  520  & $\pm$ & 50 &&   1280  & $\pm$ & 120 && 550  & $\pm$ & 40  &&  1130  & $\pm$ & 90 \\  
              Multijets &  358  & $\pm$ & 35 &&   630  & $\pm$ & 100 &&   196  & $\pm$ & 20 &&   390  & $\pm$ & 60 \\   
      $Z$+jets, diboson &  129  & $\pm$ & 14 &&   211  & $\pm$ & 19 &&   128  & $\pm$ & 12  &&  242  & $\pm$ & 20 \\   
\hline
       Total background &  2590  & $\pm$ & 60 &&  4290  & $\pm$ & 160 && 4580  & $\pm$ & 70 &&  6470  & $\pm$ & 130  \\
                 Data   &  2622  &&&&             4260  &&&&             4555  &&&&             6433  &&           \\
\hline
\end{tabular}
\end{center}  
\end{table}

\begin{table}[hbtp]\small
  \setlength{\tabcolsep}{0.4em}
\begin{center}
\caption{\label{tab:SR2tag} 
The numbers of signal and background events and the numbers of observed data events are shown in the 2-jet 2-tag and 3-jet 2-tag  signal regions.
For signal, the values correspond to expected event yields and quoted uncertainties account for the statistical uncertainty of the number of events in the simulated samples. 
The number of background events is obtained following a ML fit to the data and uncertainties contain statistical and systematic uncertainties.
}
\begin{tabular}{l S[table-number-alignment=center] p{0.05cm} S[table-column-width=3em,table-number-alignment=center] p{0.4cm} S[table-column-width=3em,table-number-alignment=center]  p{0.05cm} S[table-column-width=3em,table-number-alignment=center] p{0.4cm} S[table-column-width=3em,table-number-alignment=center] p{0.05cm} S[table-column-width=3em,table-number-alignment=center] p{0.4cm}S[table-column-width=3em,table-number-alignment=center] p{0.05cm} S[table-column-width=3em,table-column-width=3em,table-number-alignment=center] }

\hline\hline
& \multicolumn{3}{c}{2-jet 2-tag ($e^\pm$)} & & \multicolumn{3}{c}{2-jet 2-tag ($\mu^\pm$) }  & &\multicolumn{3}{c}{3-jet 2-tag ($e^\pm$)} & & \multicolumn{3}{c}{3-jet 2-tag ($\mu^\pm$)}\\
\hline                    
$W'_\text{R}$ (1.0 TeV) &     1584 & $\pm$ &      35 & &     2060 & $\pm$ &      40 & &     1241 & $\pm$ &      30 & &     1749 & $\pm$ &      34\\ 
$W'_\text{R}$ (2.0 TeV) &       33.5 & $\pm$ &       1.0 & &       55.5 & $\pm$ &       1.2 & &       51.6 & $\pm$ &       1.2 & &       84.3 & $\pm$ &       1.5\\ 
$W'_\text{R}$ (3.0 TeV) &        1.4 & $\pm$ &       0.1 & &        2.6 & $\pm$ &       0.1 & &         2.5 & $\pm$ &       0.1 & &        5.1 & $\pm$ &       0.1\\ 
$W'_\text{R}$ (4.0 TeV) &        0.131 & $\pm$ &       0.007 & &         0.25 & $\pm$ &       0.01 & &        0.21 &  $\pm$ &      0.01 & &        0.46 & $\pm$ &       0.01\\ 
$W'_\text{R}$ (5.0 TeV) &        0.035 & $\pm$ &        0.002 & &        0.053 & $\pm$ &       0.002 &  &       0.044 & $\pm$ &       0.002 & &         0.080 & $\pm$ &       0.002\\ \hline
             $t\bar{t}$ & 536  & $\pm$ & 14    && 789  & $\pm$ & 16     && 2459  & $\pm$ & 31   && 3200  & $\pm$ & 40   \\
            Single-top  &  121  & $\pm$ & 6    && 176  & $\pm$ & 10     && 235  & $\pm$ & 12    && 347  & $\pm$ & 17    \\
               $W$+jets &   28  & $\pm$ & 6    && 42  & $\pm$ & 4.0     && 50  & $\pm$ & 5      && 97  & $\pm$ & 9  \\
              Multijets &   36  & $\pm$ & 6    && 71  & $\pm$ & 13      && 95  & $\pm$ & 11     && 135  & $\pm$ & 22    \\
      $Z$+jets, diboson &  2.5  & $\pm$ & 0.4  && 11.5 & $\pm$ & 1.3    && 21.2  & $\pm$ & 2.1  && 26.9  & $\pm$ & 2.3  \\
\hline
       Total background & 723  & $\pm$ & 16    && 1088  & $\pm$ & 21   && 2859  & $\pm$ & 33   && 3810  & $\pm$ & 50   \\
                 Data   & 683  &&&&               1091  &&&&              2869  &&&&              3797  &&           \\
\hline
\end{tabular}
\end{center}  
\end{table}

\begin{figure}[hbtp]
\centering
\includegraphics[width=0.40\textwidth]{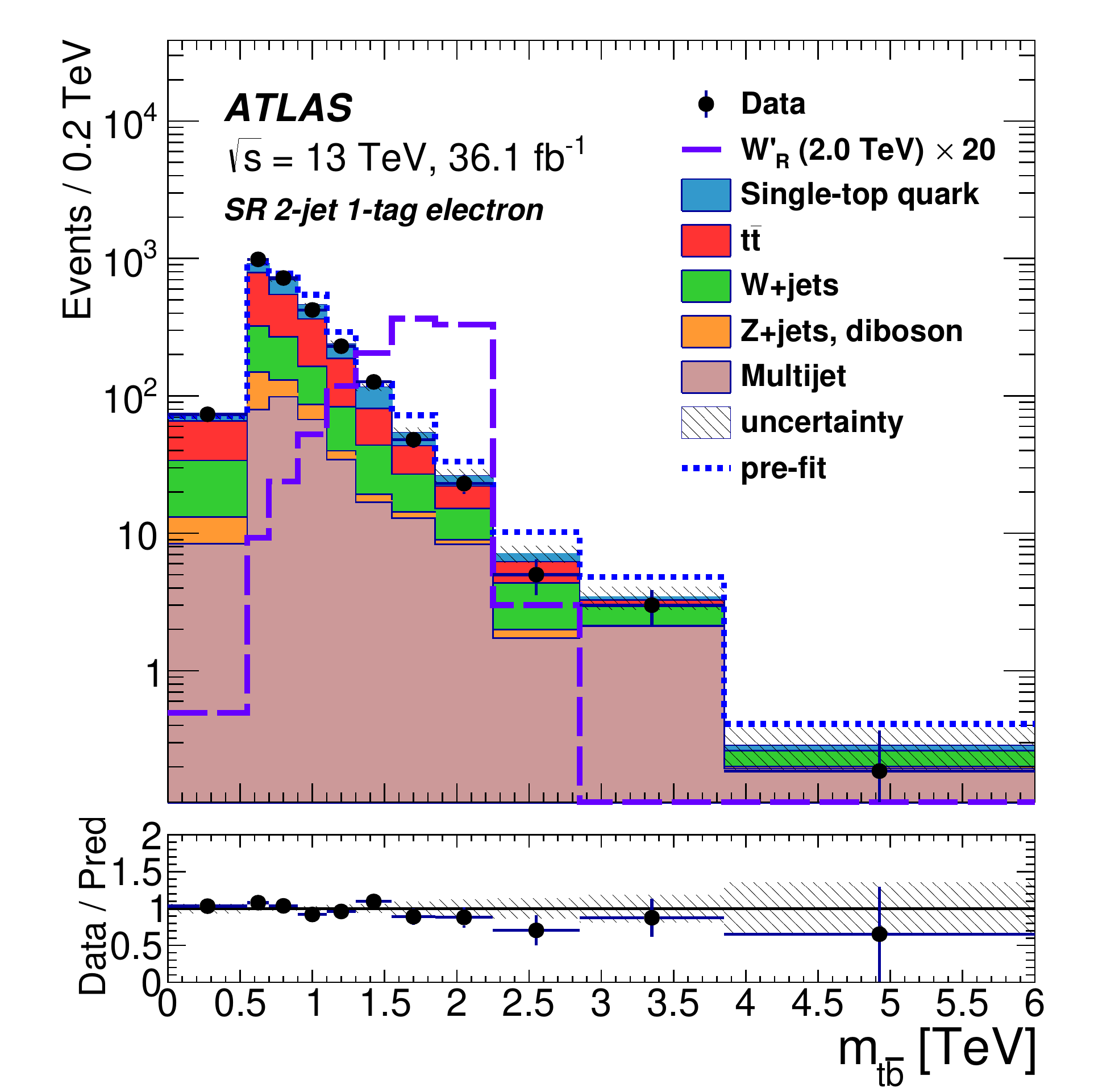}
\includegraphics[width=0.40\textwidth]{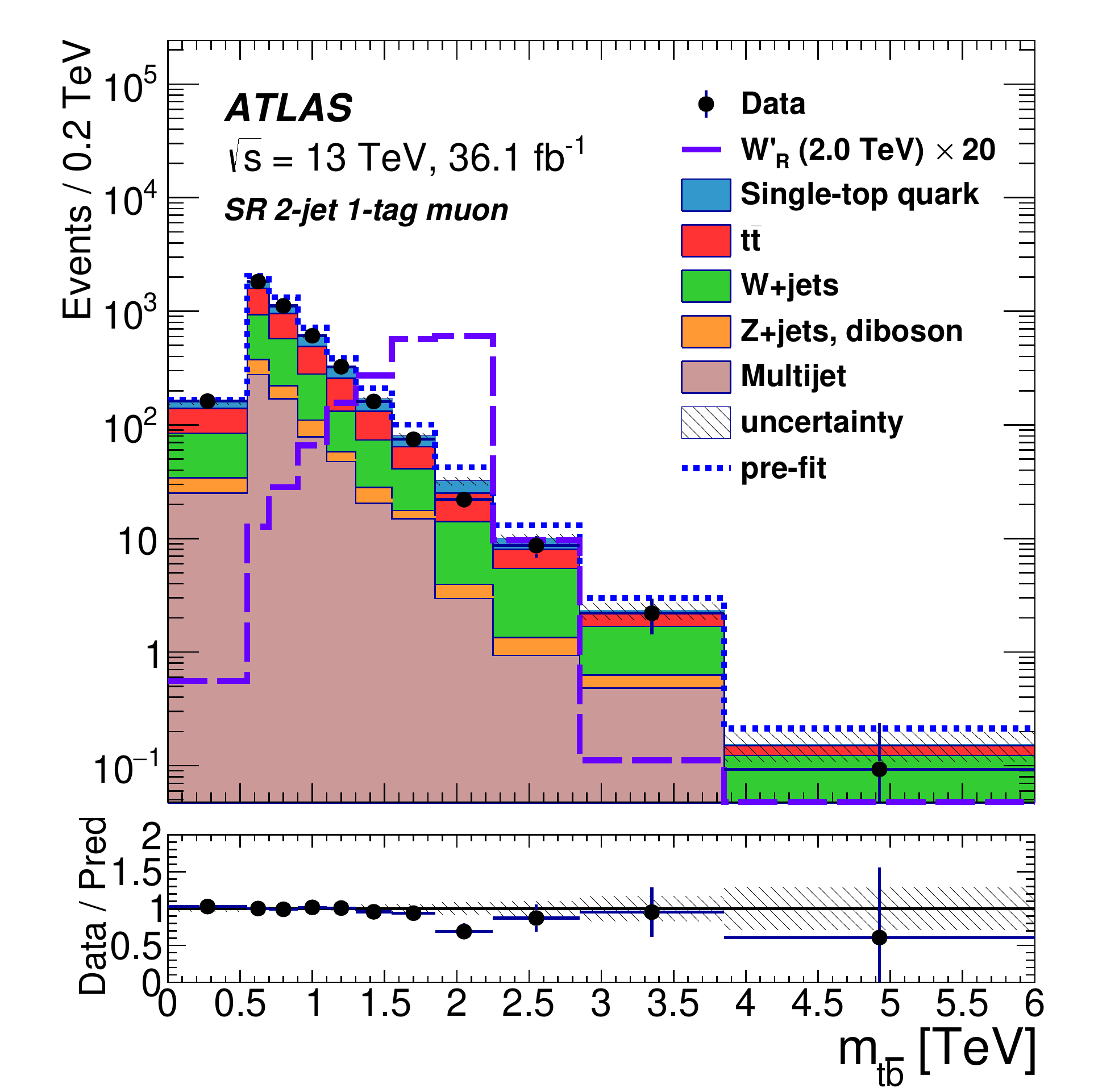}\\
\includegraphics[width=0.40\textwidth]{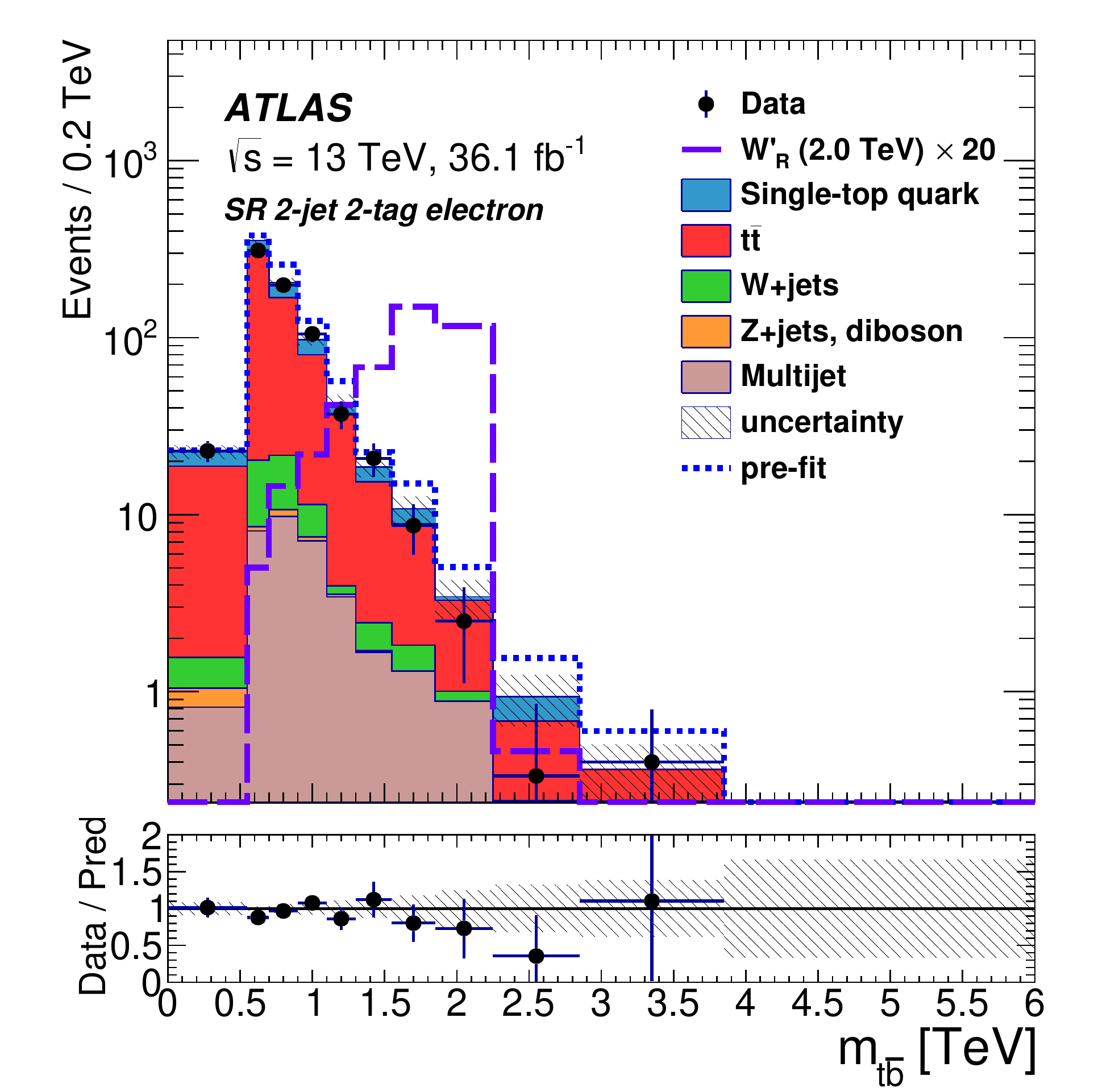}
\includegraphics[width=0.40\textwidth]{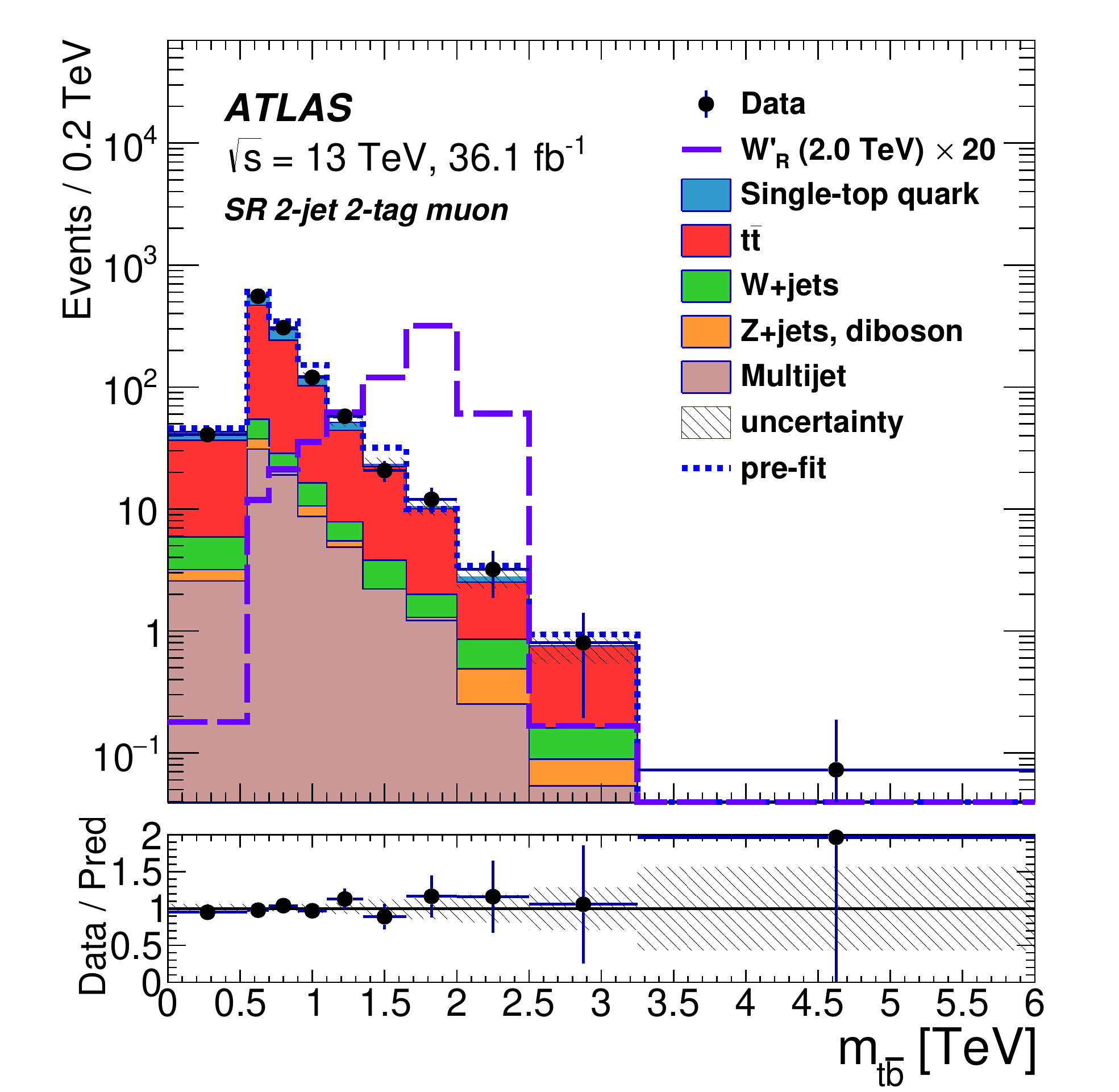}
\caption{Post-fit distributions of the reconstructed mass of the \WpR\ boson candidate in the (top) 2-jet 1-tag and (bottom) 2-jet 2-tag  signal regions, for (left) electron and (right) muon channels. 
An expected signal contribution corresponding to a \WpR\  boson mass of 2.0~\TeV\ enhanced 20 times is shown. 
The pre-fit line presents the background prediction before the fit is performed. 
Uncertainty bands include all the systematic and statistical uncertainties. The residual difference between the data and MC yields is shown as a ratio in the bottom portion of each figure, wherein the error bars on the data points correspond to the data Poisson uncertainty.}
\label{fig:SRmtb2jet}
\end{figure}

\begin{figure}[hbtp]
\centering
\includegraphics[width=0.40\textwidth]{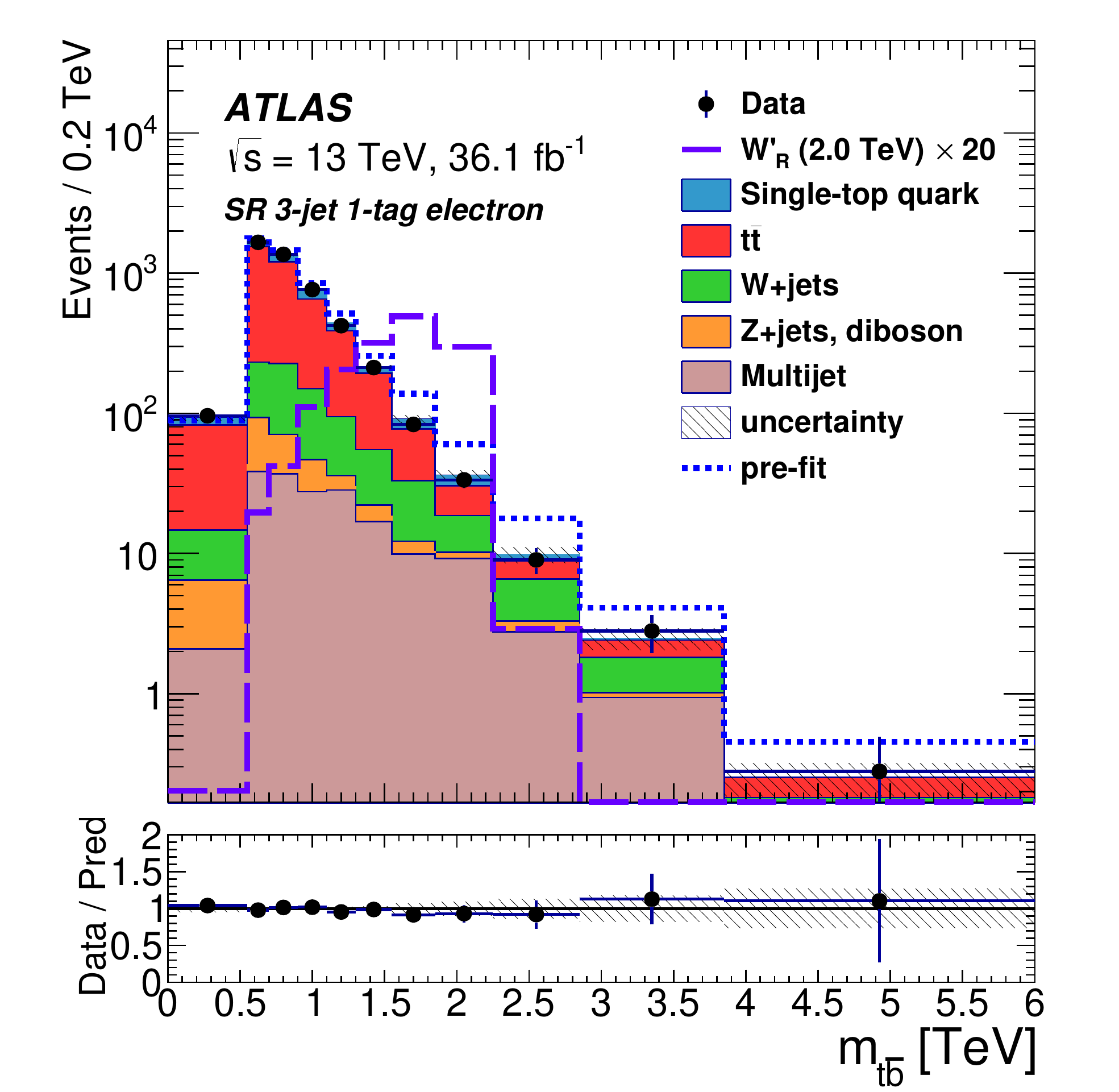}
\includegraphics[width=0.40\textwidth]{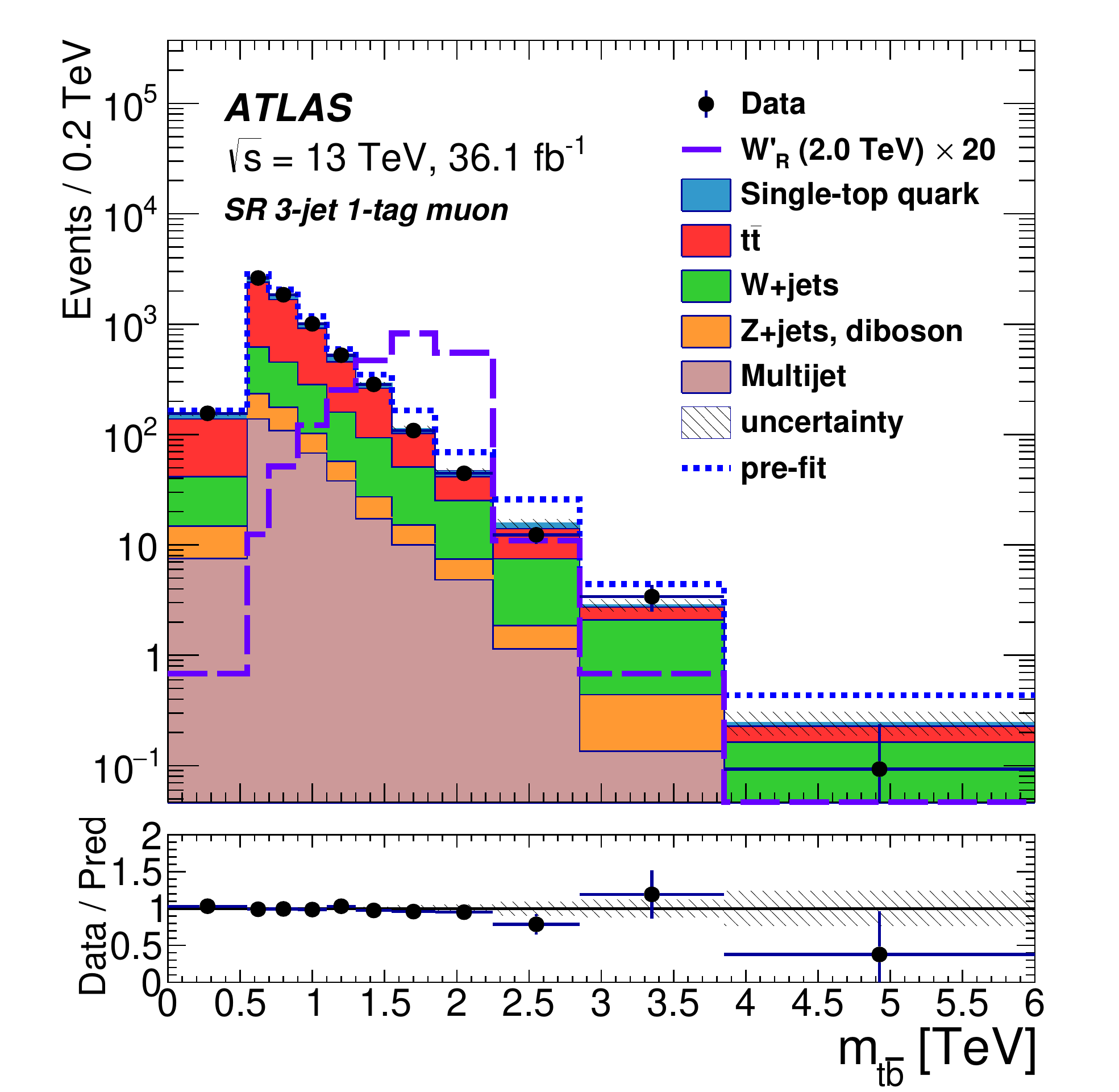}\\
\includegraphics[width=0.40\textwidth]{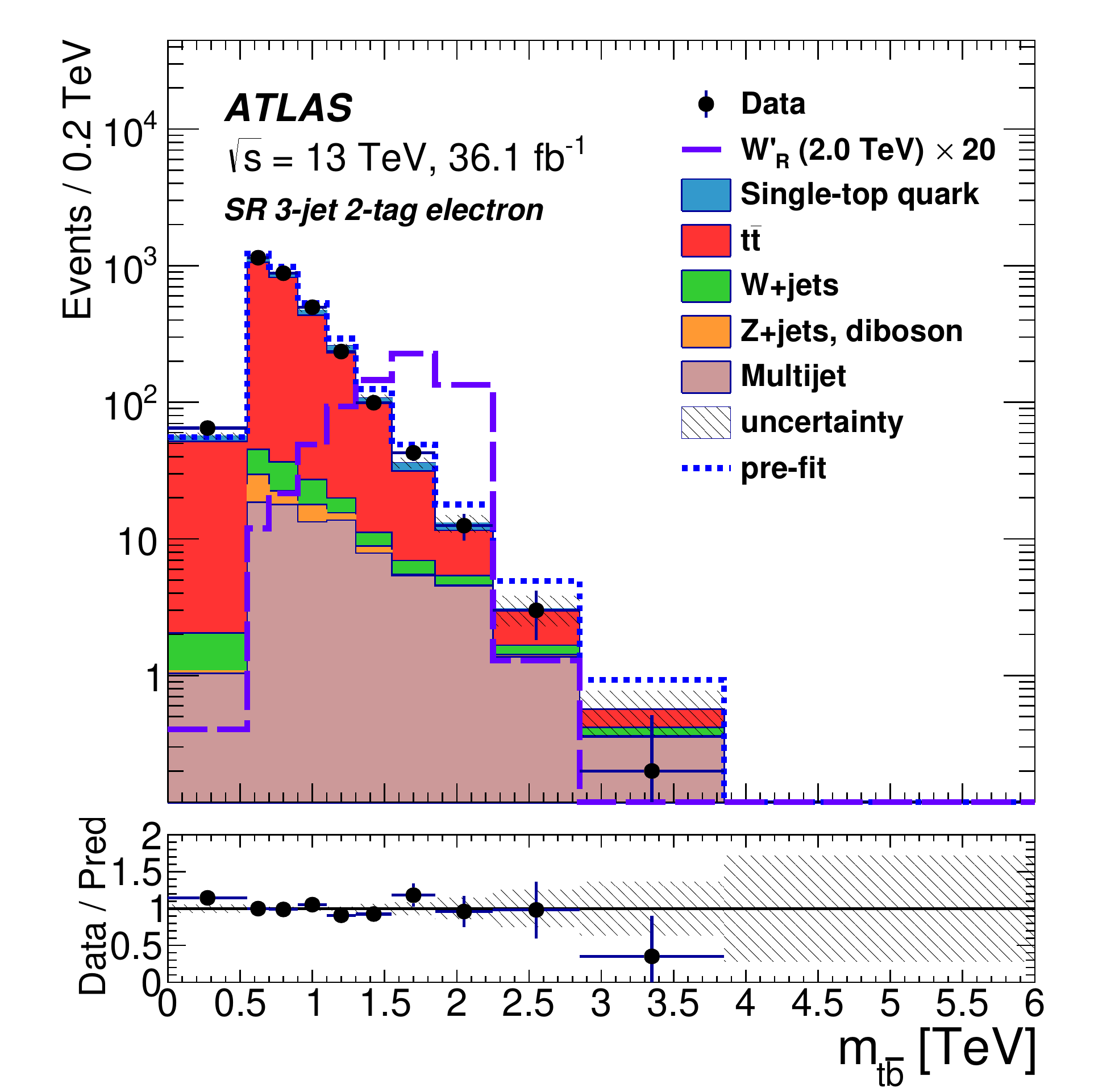}
\includegraphics[width=0.40\textwidth]{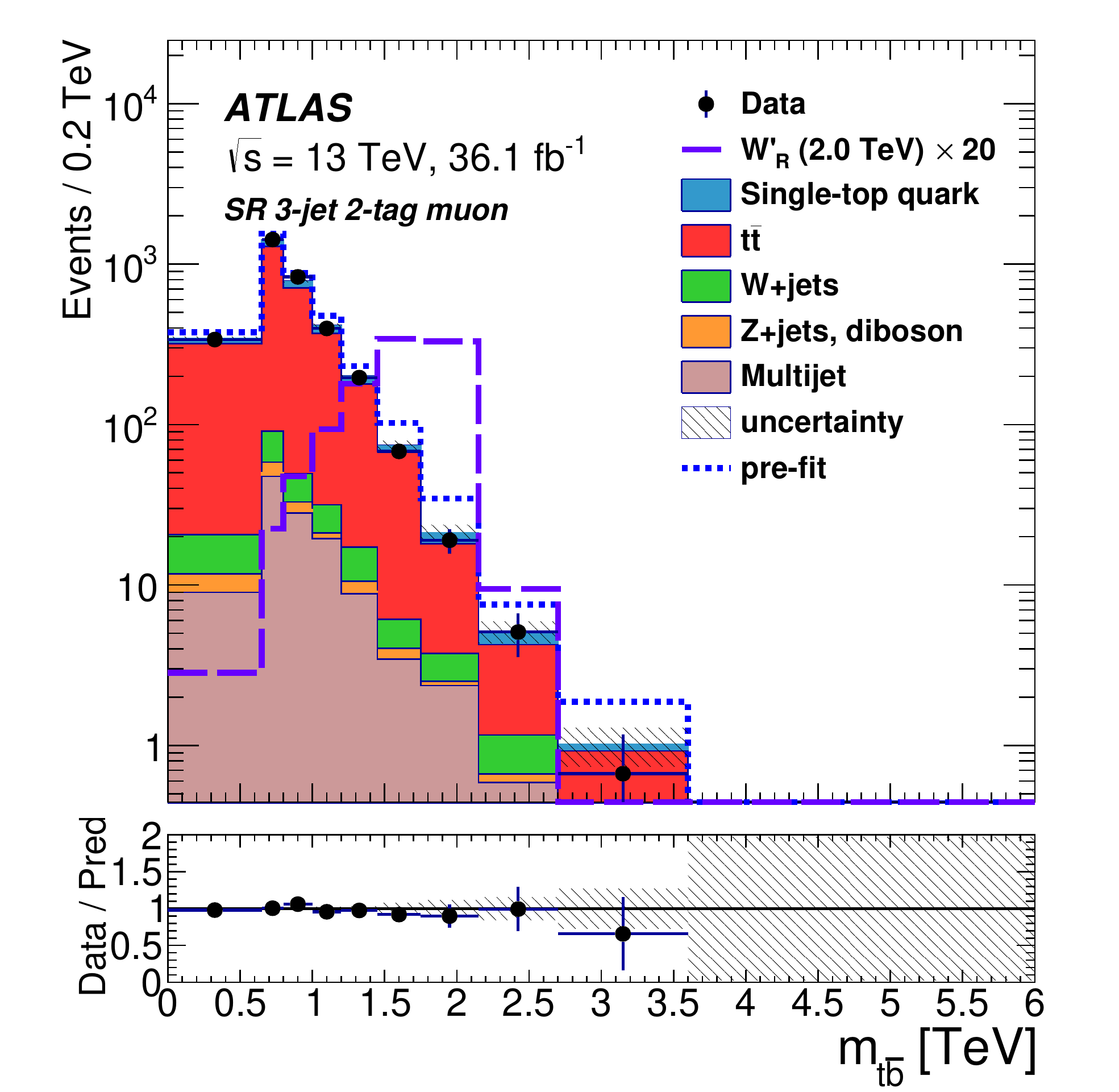}
\caption{Post-fit distributions of the reconstructed mass of the \WpR\ boson candidate in the (top) 3-jet 1-tag and (bottom) 3-jet 2-tag  signal regions, for (left) electron and (right) muon channels. 
An expected signal contribution corresponding to a \WpR\  boson mass of 2.0~\TeV\ enhanced 20 times is shown. 
The pre-fit line presents the background prediction before the fit is performed. 
Uncertainty bands include all the systematic and statistical uncertainties. The residual difference between the data and MC yields is shown as a ratio in the bottom portion of each figure, wherein the error bars on the data points correspond to the data Poisson uncertainty.}
\label{fig:SRmtb3jet}
\end{figure}

As no significant excess over the background prediction is observed, upper limits at the 95\% confidence level (CL) are set on the production cross section times the branching fraction for each model.  The limits are evaluated using a modified frequentist method known as CL$_\text{s}$~\cite{Read:2002hq} with a profile-likelihood-ratio test statistic~\cite{Cowan:2010js} using the asymptotic approximation. 

The 95\% CL upper limits on the production cross section multiplied by the branching fraction for  $\WpR \rightarrow t\bar{b}$ are shown in Figure~\ref{fig:wpLimit} as a function of the resonance mass.
The observed and expected limits are derived using a linear interpolation between simulated signal mass hypotheses.
The exclusion limits range between $4.9$~pb and $2.9\times 10^{-2}$~pb for \WpR\ boson masses from 0.5~\TeV\ to 5~\TeV. 
The lower observed limits for \WpR\ masses around 2.5~\tev\ are due to a deficit of data events in the 2--2.5~\tev\ $m_{t\bar{b}}$ range 
in the 2-jet 1-tag and 3-jet 1tag (muon) signal regions.
The existence of \WpR\ bosons with masses $m_{\WpR}$ < 3.15~\tev\  is excluded for the {\textsc ZTOP}  benchmark model for \WpR, assuming that the \WpR\ coupling $g^\prime$ is equal to the SM weak coupling constant $g$. 
 
Limits on the ratio of couplings $g'/g$ as a function of the \WpR\ boson mass can be derived from the limits on the \WpR\ boson 
cross section. Limits can also be set for $g'/g>1$, as models remain perturbative up to a ratio of about five~\cite{Duffty:2012rf}.
The \WpR\ boson cross section has a dependence on the coupling $g'$, coming from the variation of the resonance width.
The scaling of the \WpR\ boson cross section as a function of $g'/g$ and $m_{W'}$ is estimated at NLO using the {\textsc ZTOP} generator. 
In addition, specific signal samples are used in order to take into account the effect on the acceptance and on kinematical distributions of the increased signal width (compared with the nominal samples) for values of $g'/g>1$.
Figure~\ref{fig:couplingLimit} shows the excluded parameter space as a function of the \WpR\ resonance mass, wherein the effect of increasing \WpR\ width for coupling values of $g^\prime/g>1$ is included for signal acceptance and differential distributions.
The lowest observed (expected) limit on $g'/g$, obtained for a \WpR\ boson mass of 0.75~\tev, is 0.13 (0.13).

\begin{figure}[hbtp]
  \centering
  \includegraphics[width=0.7\textwidth]{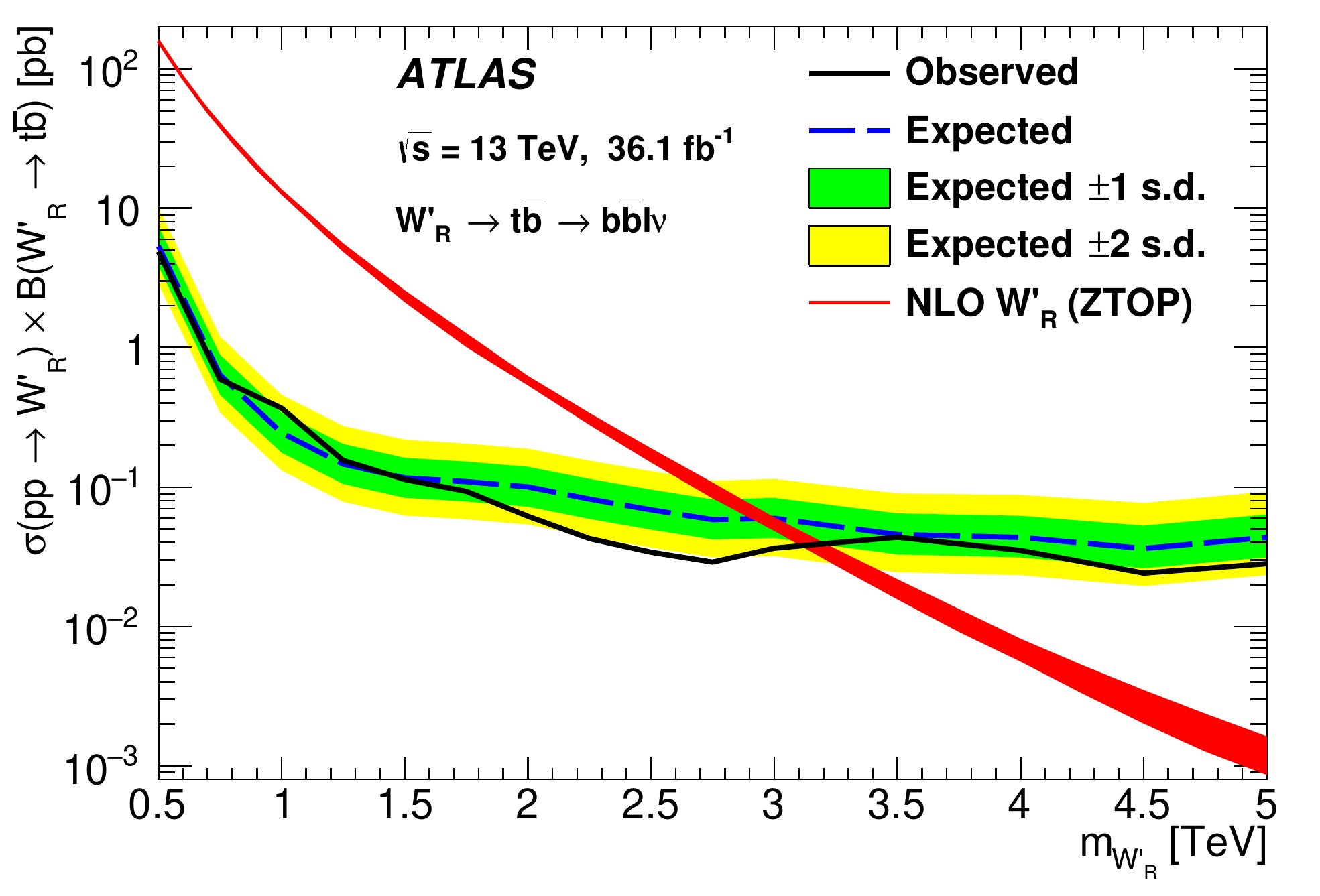}
\caption{Upper limits at the 95\% CL on the \WpR\ production cross section times the $\WpR \rightarrow t\bar{b}$ branching fraction as a function of resonance mass, assuming $g^\prime/g=1$.  The solid curve corresponds to the observed limit, while the dashed curve and shaded bands correspond to the limit expected in the absence of signal and the regions enclosing one/two standard deviation (s.d.) fluctuations of the expected limit. The prediction made by the benchmark model generator {\textsc ZTOP}~\cite{Duffty:2012rf}, and its width that correspond to variations due to scale and PDF uncertainty, are also shown.}
\label{fig:wpLimit}
\end{figure}

\begin{figure}[!h!tp]
\centering
\includegraphics[width=0.7\textwidth]{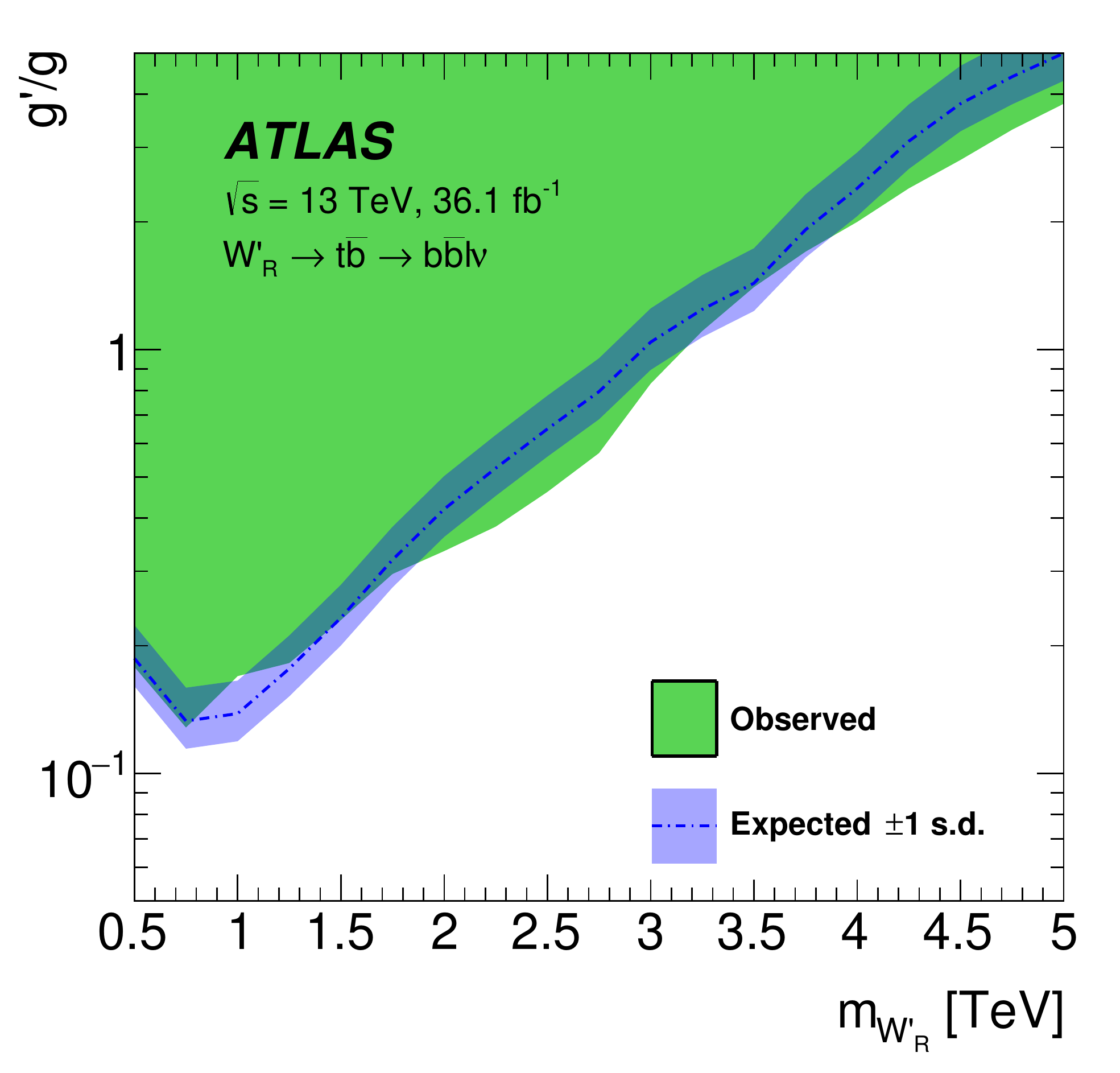}
\caption{Observed and expected 95\% CL limit on the ratio $g^\prime/g$, as a function of resonance mass, for right-handed $W^\prime$ coupling. 
The filled area correspond to the observed limit while the dashed line and the one standard deviation (s.d.) band correspond the expected limit.
The impact of the increased \WpR\  width for coupling values of $g'/g>1$ on the acceptance and on kinematical distributions is taken into account.}
\label{fig:couplingLimit}
\end{figure}

The ATLAS experiment has recently searched for $\WpR \rightarrow t\bar{b}$ in the fully hadronic final state~\cite{Aaboud:2018juj} using 36.1\ifb, corresponding to the same data collection period as the analysis presented here.  As these two searches are complementary and use mutually orthogonal event selections, a more general and powerful search for $\WpR \rightarrow t\bar{b}$ production can be obtained via their statistical combination.  The signal simulation was produced in the same manner for both searches, and the simulation of shared background sources is obtained with identical or similar tools.  The fully hadronic search has a background dominated by QCD multijet production, which is estimated via data-driven methods.  The smaller contribution from $t\bar{t}$ and singly produced top quarks is common to the two analyses, and thus all systematic uncertainties related to shared reconstruction or selection methods are treated as fully correlated.

The result of the combination of the cross section times branching fraction limits of the leptonic and fully hadronic analyses is shown in Figure~\ref{fig:combinedLimit}. 
The individual limits and their combination are shown in Figure~\ref{fig:combinedLimitComp}.
The expected limits produced by the two searches are similar above a resonance mass of 2~\TeV, below which the fully hadronic search suffers due to inefficiency from dijet trigger thresholds causing it not to contribute for resonance masses below 1~\TeV. Thus, the expected limits on the production cross section multiplied by the branching fraction improve by approximately 35\% above 1~\TeV\ and the combined result raises the lower limit on the \WpR\ mass to 3.25~\TeV. 
On the other hand, the gain from combining the observed cross section times branching fraction limits is rather modest, compared with the result of the leptonic analysis only, because of upward fluctuations observed in the fully hadronic analysis data.

\begin{figure}[!h!tp]
\centering
\includegraphics[width=0.7\textwidth]{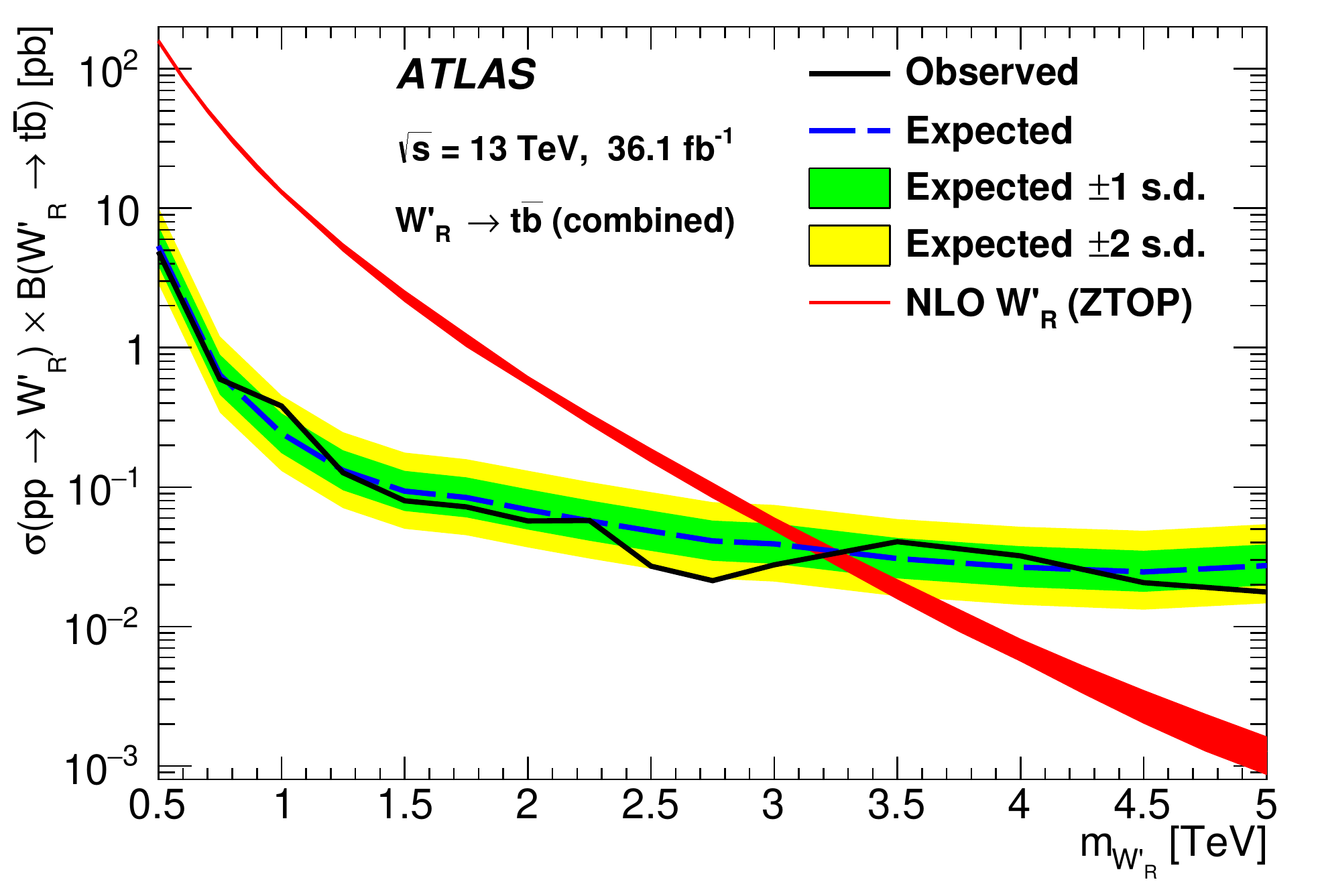}
\caption{Observed and expected 95\% CL upper limit on the \WpR\ production cross section times the $\WpR \rightarrow t\bar{b}$ branching fraction as a function of resonance mass for the combination of semileptonic and hadronic~\cite{Aaboud:2018juj} $W^\prime \rightarrow t\bar{b}$ searches, assuming $g^\prime/g=1$.  
The hadronic search covers a mass range between 1.0 and 5.0~\TeV.
The solid black curve corresponds to the observed limit, while the dashed curve and shaded bands correspond to the limit expected in the absence of signal and the regions enclosing one/two standard deviation (s.d.) fluctuations of the expected limit. The prediction made by the benchmark model generator {\textsc ZTOP}~\cite{Duffty:2012rf}, and its width that correspond to variations due to scale and PDF uncertainty, are also shown.}
\label{fig:combinedLimit}
\end{figure}

\begin{figure}[!h!tp]
\centering
\includegraphics[width=0.7\textwidth]{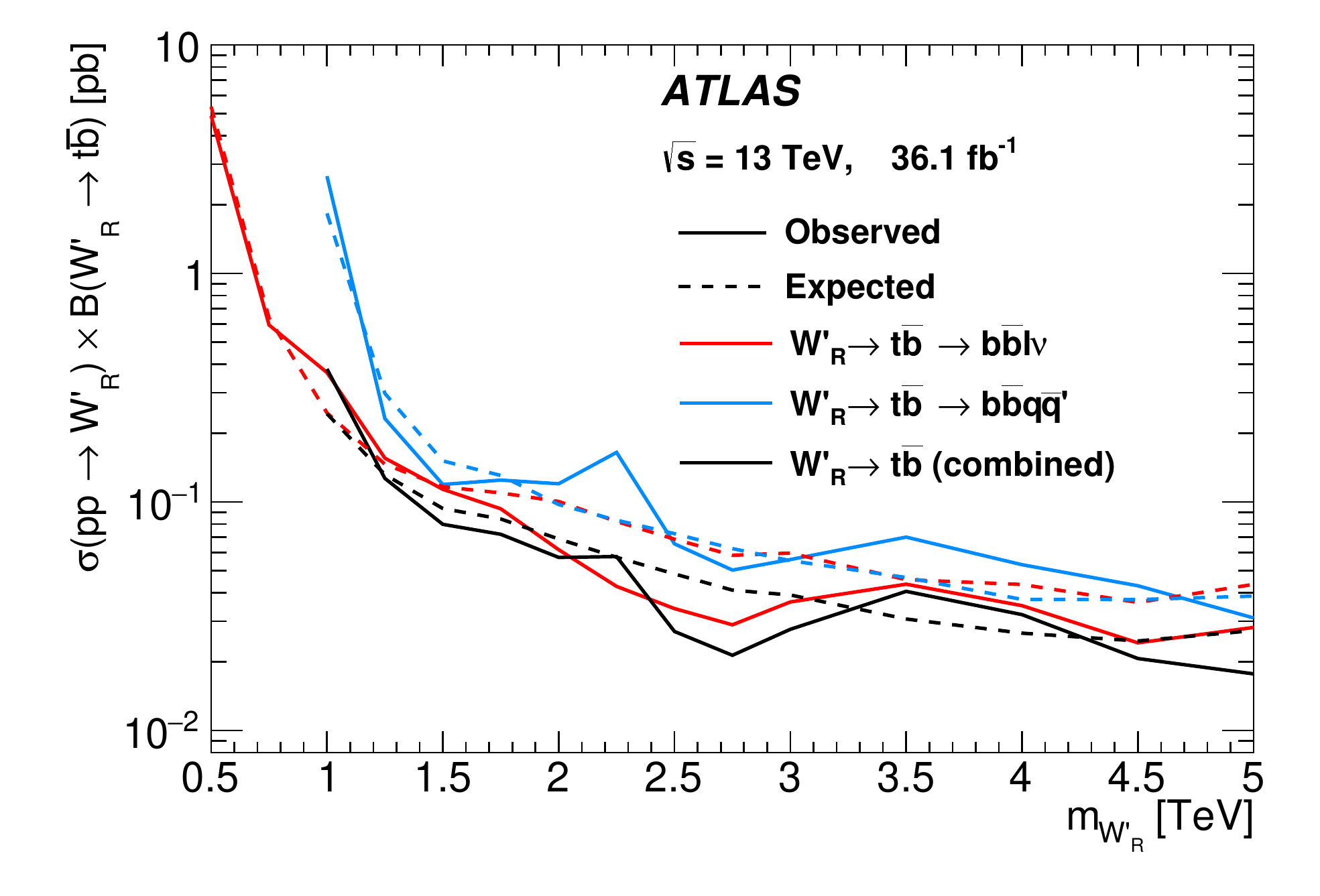}
\caption{Observed and expected 95\% CL upper limit on the \WpR\ production cross section times the $\WpR \rightarrow t\bar{b}$ branching fraction as a function of resonance mass, for the semileptonic and hadronic~\cite{Aaboud:2018juj} $W^\prime \rightarrow t\bar{b}$ searches, as well as their combination.  The solid curves correspond to the observed upper limits, while the dashed lines are the expected limits.}
\label{fig:combinedLimitComp}
\end{figure}

%-------------------------------------------------------------------------------
\clearpage
\section{Conclusion}
\label{sec:conclusion}

A search for $\WpR \to t\bar{b}$ in the  lepton plus jets final state is performed using \lumi\ of 13~\TeV\ $pp$ collision data collected with the ATLAS detector at the LHC. No significant excess of events is observed above the SM predictions.
Upper limits are placed at the 95\% CL on the cross section times branching fraction, $\sigma(pp\rightarrow \WpR \rightarrow t\bar{b}$), ranging between $4.9$~pb and $2.9\times 10^{-2}$~pb in the mass range  of 0.5~\TeV\ to 5~\TeV\ for a right-handed \Wp\ boson.
Exclusion limits are also calculated for the ratio of the couplings $g'/g$ and the lowest observed limit, obtained for a \WpR\ boson mass of 0.75~\tev, is 0.13.
A statistical combination of the cross-section limits is performed with the results obtained when the fully hadronic decays of $\WpR\ \rightarrow t\bar{b}$ are considered.
The upper limits on the cross section times branching fraction improve by approximately 35\% above 1~\TeV.
Masses below 3.15 (3.25)~\TeV\ are excluded for \WpR\ bosons in the benchmark {\textsc ZTOP} model for the semileptonic (combined semileptonic and hadronic) scenarios.

%-------------------------------------------------------------------------------
\section*{Acknowledgements}
%-------------------------------------------------------------------------------
% Acknowledgements for papers with collision data
% Version 05-Oct-2018

% Standard acknowledgements start here
%----------------------------------------------
We thank CERN for the very successful operation of the LHC, as well as the
support staff from our institutions without whom ATLAS could not be
operated efficiently.

We acknowledge the support of ANPCyT, Argentina; YerPhI, Armenia; ARC, Australia; BMWFW and FWF, Austria; ANAS, Azerbaijan; SSTC, Belarus; CNPq and FAPESP, Brazil; NSERC, NRC and CFI, Canada; CERN; CONICYT, Chile; CAS, MOST and NSFC, China; COLCIENCIAS, Colombia; MSMT CR, MPO CR and VSC CR, Czech Republic; DNRF and DNSRC, Denmark; IN2P3-CNRS, CEA-DRF/IRFU, France; SRNSFG, Georgia; BMBF, HGF, and MPG, Germany; GSRT, Greece; RGC, Hong Kong SAR, China; ISF and Benoziyo Center, Israel; INFN, Italy; MEXT and JSPS, Japan; CNRST, Morocco; NWO, Netherlands; RCN, Norway; MNiSW and NCN, Poland; FCT, Portugal; MNE/IFA, Romania; MES of Russia and NRC KI, Russian Federation; JINR; MESTD, Serbia; MSSR, Slovakia; ARRS and MIZ\v{S}, Slovenia; DST/NRF, South Africa; MINECO, Spain; SRC and Wallenberg Foundation, Sweden; SERI, SNSF and Cantons of Bern and Geneva, Switzerland; MOST, Taiwan; TAEK, Turkey; STFC, United Kingdom; DOE and NSF, United States of America. In addition, individual groups and members have received support from BCKDF, the Canada Council, CANARIE, CRC, Compute Canada, FQRNT, and the Ontario Innovation Trust, Canada; EPLANET, ERC, ERDF, FP7, Horizon 2020 and Marie Sk{\l}odowska-Curie Actions, European Union; Investissements d'Avenir Labex and Idex, ANR, R{\'e}gion Auvergne and Fondation Partager le Savoir, France; DFG and AvH Foundation, Germany; Herakleitos, Thales and Aristeia programmes co-financed by EU-ESF and the Greek NSRF; BSF, GIF and Minerva, Israel; BRF, Norway; CERCA Programme Generalitat de Catalunya, Generalitat Valenciana, Spain; the Royal Society and Leverhulme Trust, United Kingdom.

The crucial computing support from all WLCG partners is acknowledged gratefully, in particular from CERN, the ATLAS Tier-1 facilities at TRIUMF (Canada), NDGF (Denmark, Norway, Sweden), CC-IN2P3 (France), KIT/GridKA (Germany), INFN-CNAF (Italy), NL-T1 (Netherlands), PIC (Spain), ASGC (Taiwan), RAL (UK) and BNL (USA), the Tier-2 facilities worldwide and large non-WLCG resource providers. Major contributors of computing resources are listed in Ref.~\cite{ATL-GEN-PUB-2016-002}.
%----------------------------------------------

%-------------------------------------------------------------------------------
% If you use biblatex and etextither biber or bibtex to process the bibliography
% just say \printbibliography here
\printbibliography
% If you want to use the tradtextitional BibTeX you need to use the syntax below.
%\bibliographystyle{bibtex/bst/atlasBibStyleWoTitle}
%\bibliography{paper,bibtex/bib/ATLAS}
%-------------------------------------------------------------------------------
%-------------------------------------------------------------------------------
% Print the list of contributors to the analysis
% The argument gives the fraction of the text width used for the names
%-------------------------------------------------------------------------------
\clearpage
% ATLAS Collaboration author list
% Reference date of EXOT-2016-18 is 2018-04-24
% Author list last updated on date 23-JUL-18
% Data extracted on 23-Jul-2018 for paper reference EXOT-2016-18
% at 8:37am
 
\begin{flushleft}
{\Large The ATLAS Collaboration}

\bigskip

M.~Aaboud$^\textrm{\scriptsize 34d}$,    
G.~Aad$^\textrm{\scriptsize 99}$,    
B.~Abbott$^\textrm{\scriptsize 124}$,    
O.~Abdinov$^\textrm{\scriptsize 13,*}$,    
B.~Abeloos$^\textrm{\scriptsize 128}$,    
D.K.~Abhayasinghe$^\textrm{\scriptsize 91}$,    
S.H.~Abidi$^\textrm{\scriptsize 164}$,    
O.S.~AbouZeid$^\textrm{\scriptsize 39}$,    
N.L.~Abraham$^\textrm{\scriptsize 153}$,    
H.~Abramowicz$^\textrm{\scriptsize 158}$
% ...
\bigskip
\\

% ...
$^{13}$Institute of Physics, Azerbaijan Academy of Sciences, Baku; Azerbaijan.\\
% ...
$^{34}$$^{(a)}$Facult\'e des Sciences Ain Chock, R\'eseau Universitaire de Physique des Hautes Energies - Universit\'e Hassan II, Casablanca;$^{(b)}$Centre National de l'Energie des Sciences Techniques Nucleaires (CNESTEN), Rabat;$^{(c)}$Facult\'e des Sciences Semlalia, Universit\'e Cadi Ayyad, LPHEA-Marrakech;$^{(d)}$Facult\'e des Sciences, Universit\'e Mohamed Premier and LPTPM, Oujda;$^{(e)}$Facult\'e des sciences, Universit\'e Mohammed V, Rabat; Morocco.\\
% ...
$^{39}$Niels Bohr Institute, University of Copenhagen, Copenhagen; Denmark.\\
% ...
$^{91}$Department of Physics, Royal Holloway University of London, Egham; United Kingdom.\\
% ...
$^{99}$CPPM, Aix-Marseille Universit\'e, CNRS/IN2P3, Marseille; France.\\
% ...
$^{124}$Homer L. Dodge Department of Physics and Astronomy, University of Oklahoma, Norman OK; United States of America.\\
% ...
$^{128}$LAL, Universit\'e Paris-Sud, CNRS/IN2P3, Universit\'e Paris-Saclay, Orsay; France.\\
% ...
$^{153}$Department of Physics and Astronomy, University of Sussex, Brighton; United Kingdom.\\
% ...
$^{158}$Raymond and Beverly Sackler School of Physics and Astronomy, Tel Aviv University, Tel Aviv; Israel.\\
% ...
$^{164}$Department of Physics, University of Toronto, Toronto ON; Canada.\\
% ...
$^{a}$ Also at Borough of Manhattan Community College, City University of New York, NY; United States of America.\\
$^{b}$ Also at Centre for High Performance Computing, CSIR Campus, Rosebank, Cape Town; South Africa.\\
$^{c}$ Also at CERN, Geneva; Switzerland.\\
$^{d}$ Also at CPPM, Aix-Marseille Universit\'e, CNRS/IN2P3, Marseille; France.\\
$^{e}$ Also at D\'epartement de Physique Nucl\'eaire et Corpusculaire, Universit\'e de Gen\`eve, Gen\`eve; Switzerland.\\
% ...
$^{*}$ Deceased

\end{flushleft}

% Created with Glance <Atlas.Glance@cern.ch>

%\PrintAtlasContribute{0.30}
%-------------------------------------------------------------------------------
%\clearpage
%\part*{Auxiliary material}
%\label{sec:auxiliary}
%\addcontentsline{toc}{part}{Auxiliary material}
%\input{EXOT-2016-18-PAPER-auxmat}
%-------------------------------------------------------------------------------
\end{document}